\documentclass[prb,twocolumn,superscriptaddress,floats,citeautoscript]{revtex4}

\usepackage{graphicx}
\usepackage{amsmath}
\usepackage{dcolumn}
\usepackage{times}
\usepackage{bm}
\graphicspath{{.}{./EPS/}}

\begin{document}

\title{Spin-3/2 physics of semiconductor hole nanowires: Valence-band mixing and 
tunable interplay between bulk-material and orbital bound-state spin splittings}

\author{D. Csontos}
\affiliation{Institute of Fundamental Sciences and MacDiarmid Institute for Advanced 
Materials and Nanotechnology, Massey University (Manawatu Campus), Private Bag
11~222, Palmerston North, New Zealand}

\author{P. Brusheim}
\affiliation{Division of Solid State Physics, Lund University, Box 118, S-22100 Lund, 
Sweden}
\affiliation{Institute of High Performance Computing, 1 Fusionopolis Way, \#16-16
Connexis, Singapore 138632, Singapore}

\author{U. Z\"ulicke}
\affiliation{Institute of Fundamental Sciences and MacDiarmid Institute for Advanced 
Materials and Nanotechnology, Massey University (Manawatu Campus), Private Bag
11~222, Palmerston North, New Zealand}
\affiliation{Centre for Theoretical Chemistry and Physics, Massey University (Albany
Campus), Private Bag 102904, North Shore MSC, Auckland 0745, New Zealand}

\author{H. Q. Xu}
\affiliation{Division of Solid State Physics, Lund University, Box 118, S-22100 Lund, 
Sweden}

\date{\today}

\begin{abstract}
We present a detailed theoretical study of the electronic spectrum and Zeeman
splitting in hole quantum wires. The spin-3/2 character of the topmost
bulk-valence-band states results in a strong variation of subband-edge $g$ factors
between different subbands. We elucidate the interplay between quantum 
confinement and heavy-hole~--~light-hole mixing and identify a certain robustness
displayed by low-lying hole-wire subband edges with respect to changes in the
shape or strength of the wire potential. The ability to address individual subband
edges in, e.g., transport or optical experiments enables the study of holes states with
nonstandard spin polarization, which do not exist in spin-1/2 systems. Changing the
aspect ratio of hole wires with rectangular cross-section turns out to strongly
affect the $g$ factor of subband edges, providing an opportunity for versatile in-situ
tuning of hole-spin properties with possible application in spintronics. The relative
importance of cubic crystal symmetry is discussed, as well as the spin splitting away
from zone-center subband edges. 
\end{abstract}

\maketitle
\section{Introduction}

Self-assembled semiconductor nanowires have attracted a lot of attention recently 
due to their well-defined crystalline structure, unique electrical and optical properties,
as well as a promising outlook for their use as building blocks in nanoelectronics, 
nanospintronics and nanobiotechnology~\cite{hirumaJAP1995,gudiksenJPCB2001,
duanNATURE2001,TangSCIENCE2002,katzPRL2002,johnsonNATMAT2002,
bjorkNANOLETT2002,bjorkAPL2002,gudiksenNATURE2002,
krishnamachariAPL2004,samuelsonPHYSICAE2004,huangSMALL2005,
thelanderNANOLETT2005,yangSCIENCE2005,luJPD2006,
pauzauskieMATERIALSTODAY2006,perssonNANOLETT2006,
samuelsonNANOTECH2006,johanssonNATMAT2006,janikAPL2006,
thelanderMATERIALSTODAY2006,martelliNANOLETT2006,lieberNATURE2006,
fengNANOLETT2007,NeretinaNANOTECH2007,liNANOLETT2007,
zhuNANOLETT2007,hochbaumNATURE2008,sorensenAPL2008,
jagadishNANOTECH2008,roddaroPRL2008,wojtowiczNANOLETT2008}.
Rapid developments in material science and technology have enabled the 
realization of nanowires from many materials and materials combinations, such as 
III-V semiconductors (e.g., GaAs~\cite{hirumaJAP1995, martelliNANOLETT2006, 
zhuNANOLETT2007}, InP~\cite{gudiksenJPCB2001, duanNATURE2001, 
krishnamachariAPL2004, huangSMALL2005}, InAs~\cite{hirumaJAP1995, 
bjorkNANOLETT2002, bjorkAPL2002, sorensenAPL2008}\/), II-VI semiconductors 
(e.g., CdSe~\cite{katzPRL2002, huangSMALL2005}, CdS~\cite{huangSMALL2005}, 
CdTe~\cite{TangSCIENCE2002,NeretinaNANOTECH2007}, ZnTe~\cite
{janikAPL2006,wojtowiczNANOLETT2008}\/), as well as Si~\cite
{yangSCIENCE2005, fengNANOLETT2007, hochbaumNATURE2008} and Ge~\cite
{lieberNATURE2006,jagadishNANOTECH2008,roddaroPRL2008}. By now it is
possible to dope nanowires~\cite{martelliNANOLETT2006,sorensenAPL2008,
liNANOLETT2007} and perform heterostructure engineering with atomic
precision~\cite{bjorkNANOLETT2002, bjorkAPL2002,huangSMALL2005,
perssonNANOLETT2006}. This enabled the realization of {\it p}-type~\cite
{martelliNANOLETT2006, liNANOLETT2007} or ambipolar~\cite{sorensenAPL2008}
transport characteristics, as well as the fabrication of nanostructures (quantum
dots~\cite{bjorkAPL2002}, rods~\cite{janikAPL2006}, and 
superlattices~\cite{gudiksenNATURE2002}\/). Based on these material-science 
developments, devices such as nanowire diodes~\cite{hirumaJAP1995,
duanNATURE2001}, superlattices for nanoscale photonics and electronics~\cite
{gudiksenNATURE2002}, lasers~\cite{johnsonNATMAT2002, 
pauzauskieMATERIALSTODAY2006}, resonant tunneling diodes~\cite
{bjorkAPL2002}, and light-emitting diodes~\cite{hirumaJAP1995, 
duanNATURE2001,huangSMALL2005}, have been demonstrated.

Another area of significant research that has emerged in recent years focuses on the study of
the spin degree of freedom. This is relevant for the fundamental understanding of spin 
phenomena in solid-state systems, but also for potential applications that utilize the spin,
rather than the charge degree of freedom, in
electronics~\cite{wolfSCIENCE2001,zuticRMP2004}. 
Spin injection, manipulation and detection has been achieved with the use of magnetic fields
and diluted magnetic semiconductors~\cite{molenkampNATURE1999,ohnoNATURE1999}. A 
complementary direction of current spin-electronics research is based on the
quantum-mechanical coupling between spin and orbital degrees of freedom. For 
example, the tunability of structural inversion-asymmetry induced spin-orbit (SO)
coupling~\cite{bychkovJPC1984} could be used to manipulate spin-polarized
currents in ballistic mesoscopic channels~\cite{dattaAPL1990}.

To study SO effects in semiconductors, charge carriers from the valence band (i.e.,
holes) are particularly interesting since they are subject to an inherently strong SO
coupling even in the bulk. In addition, as valence-band states are predominantly $p$-like
(i.e., have orbital angular momentum $L=1$), they are characterized by a total angular momentum of either $J=3/2$ or $J=1/2$. The topmost valence bands are 
described by the former; thus they behave like spin-3/2 particles and, when confined 
in nanostructures, exhibit quite counterintuitive quantum effects. For example, the
spin-3/2 nature of the topmost valence bands leads to large anisotropies of the hole $g$
factor in quantum wells~\cite{winklerbook,winklerPRL2000}, wires~\cite{zhuNANOLETT2007}, 
point contacts~\cite{danneauPRL2006, koduvayurPRL2008}, 
quantum dots\cite{pradoPRB2004,pryorPRL2006,zhangAPL2007}, and localized
acceptor states~\cite{haendelPRL2006}.

\begin{figure}[b]
\includegraphics[width=2in]{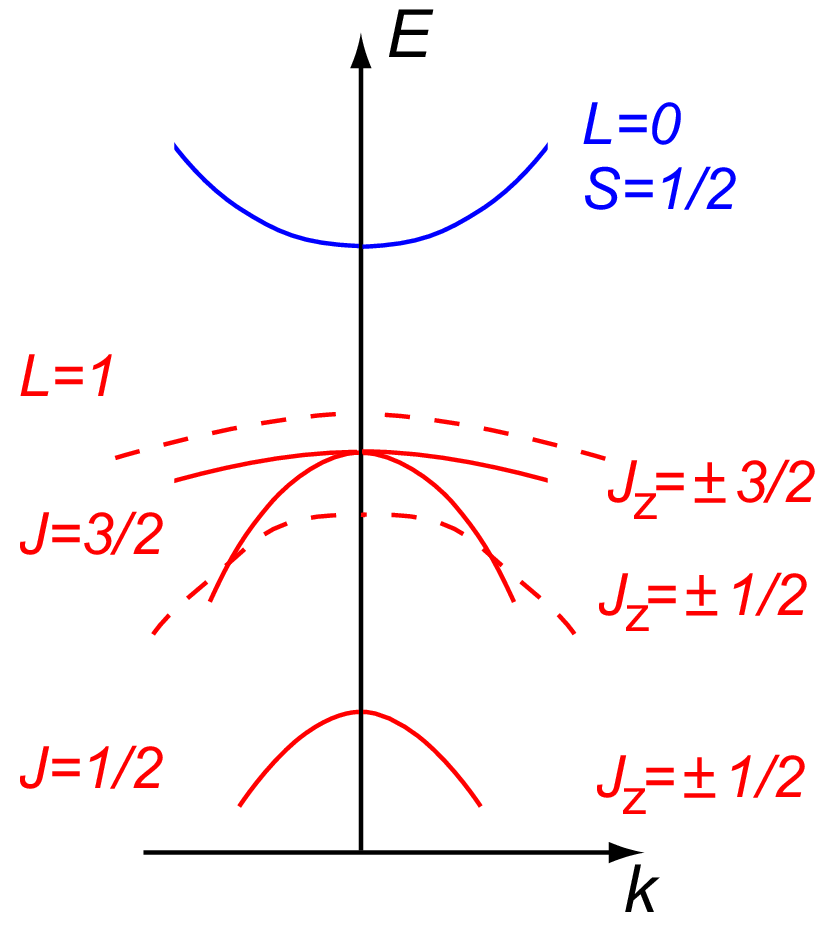} 
\caption{(Color online) Schematic dispersion of bulk-semiconductor conduction and
valence bands (solid lines). The dashed lines depict a situation where the $J=3/2$
valence bands are split due to quantum confinement, e.g., in a two-dimensional hole
gas.  \label{fig:disp}}
\end{figure}
The bulk dispersions for the conduction and valence bands of a typical semiconductor in
zero magnetic field are shown as the solid curves in Fig.~\ref{fig:disp}. The four-fold
degeneracy of the $J=3/2$ valence-band edges is lifted at finite wave vector
${\mathbf k}$~\cite{luttingerPR1956}, giving rise to separated heavy-hole (HH) and light-hole
(LH) branches whose states are distinguished by the $z$-axis spin-3/2 projection 
quantum numbers $J_{z}=\pm 3/2$ and $J_{z}=\pm 1/2$, respectively. Size 
quantization, e.g., due to a quantum-well confinement, causes an energy splitting
between the quasi-twodimensional (2D) HH and 2D LH band
edges~\cite{suzukiPRB1974,shermanPLA88,winklerbook}. This HH-LH {\em splitting}
is illustrated by the dashed lines in Fig.~\ref{fig:disp}. At finite wave vector
${\mathbf k}_{\perp}$ for motion perpendicular to the quantum-well growth direction, a
HH-LH {\em mixing} occurs that can give rise to level anticrossings~\cite{winklerbook},
further increasing the complexity of holes' electronic properties. 

While in 2D systems the HH-LH mixing is absent at the subband edge and can be 
assumed to have a small magnitude for finite $k_{\perp}$ in the presence of strong 
quantum confinement~\cite{luPRB2006}, this is not the case for hole nanowires~\cite
{bastardrev,perssonNANOLETT2004,zhangEPJB2006,haradaPRB2006,
zhuNANOLETT2007,csontosPRB2007} and quantum dots~\cite
{kyrchenkoPRB2004,lewyanvoonNANOLETT2004, shengPRB2007, 
troncaleAPL2007}. Previous studies of the topmost hole-nanowire eigenstates have 
shown that they are, to a varying degree, mixtures of HH and LH states~\cite
{csontosPRB2007,csontosAPL2008}. This has been found to influence, e.g., the spin 
splitting in cylindrical hole nanowires with hard-wall confinement\cite
{csontosPRB2007,csontosAPL2008,csontosPRB2008} and nanosized clusters.\cite
{chenPRB2005}

In this article, we present an extensive theoretical study of the electronic structure of
hole nanowires, focusing especially on the importance of HH-LH mixing and splitting. Furthermore, we have investigated the ramifications of the spin-3/2 nature of holes
for the spin splitting of hole nanowire states, in particular elucidating the interplay between the bulk-material and bound-state orbital effects due to a magnetic field. 

We use the Luttinger description~\cite{luttingerPR1956} of the semiconductor valence 
band to study the subband edges ($k_{z}=0$) of hole nanowires oriented along the
$z$ direction. In order to quantitatively analyze the strength of HH-LH mixing at the 
subband edge states, we make use of scalar invariants obtainable from a 
decomposition of the spin-3/2 density matrix in terms of a multipole expansion~\cite
{winklerPRB2004}. The squared dipole moment, which apart from a prefactor is a 
measure of the squared magnitude of the hole-spin polarization, is by its definition
\cite{winklerPRB2004} independent of the quantization axis of the angular momentum 
used in the basis functions of the hole subband edge state. Thus, it provides a 
rigorous, basis-independent way of characterizing the hole-spin polarization character 
of hole subband edge states.

We present a comprehensive analysis of the hole-spin polarization character of hole 
subband edge states in nanowires with different confining potentials. The strong
HH-LH mixing manifests as large spatial variations in the hole-spin polarization 
density profile of individual subband states. Furthermore, we observe large 
differences between the hole-spin polarization density profiles of different subband 
states. We also find that the topmost hole subband edges in general display 
universal features that are robust against changes of the confining potential's 
symmetry. For hole subband edge states that are far away from the bulk valence 
band edge, however, differences in the strength and shape of the confining potential 
significantly affect the wavefunction's hole-spin polarization.

A direct manifestation of the strong HH-LH mixing (and splitting) in hole nanowires is 
observed in the spin splitting due to a finite magnetic field. We use perturbative and 
numerical calculations to show that, in a magnetic field applied parallel to the wire 
axis, the effective $g$ factor varies strongly between different wire subband edges.
We analyze the separate contributions to the $g$ factor arising from bulk-material
and orbital bound-state interactions with the magnetic field. We find that the hole
envelope functions acquire large orbital momentum due to the imposed quantum
confinement, which strongly affects the effective $g$ factor of hole nanowire states.
Interestingly, the bulk-material and bound-state contributions can come with the same
or opposite sign, thus effectively enhancing or suppressing each other's contribution
to the total spin splitting. We also find that the interplay between the two contributions
is very sensitive to the hole wave function, and thus, that the $g$ factor of individual
wire states can be tuned by changing the confining potential. For example, we show
that by changing the aspect ratio of the lateral dimensions of a wire with rectangular
crossection, one can tune the magnitude and even the sign of the $g$ factor quite
sensitively. Such a situation can, in principle, be easily realized by tuning the lateral
confinement via side gates in self-assembled\cite{ngNANOLETT2004,
bryllertEDL2006} or lithographically defined nanowires\cite{wangAPL2000}, or
quantum point
contacts\cite{danneauAPL2006,klochanAPL2006,ensslinAIPPROC2007}.
Quasi-onedimensional (1D) hole systems thus provide an interesting laboratory for the 
study of spin-3/2 physics, as well as have the potential for being building blocks in 
hole-spintronics applications.

In the following, we start by presenting the basic theoretical formalism for describing
hole-nanowire states in the absence of a magnetic field within the Luttinger model 
(Sec.~\ref{theory}). Subsequently, we will discuss the influence of valence-band
mixing on hole-nanowire subband edges ($k_{z}=0$) in Sec.~\ref{bandmixing}, using
scalar invariants from the decomposition of the spin-3/2 density matrix to provide a
comprehensive analysis of the hole-spin polarization of nanowire subband edges. A
discussion of the influence of symmetry and strength of the confining potential will be
given. Next, the spin splitting of hole-nanowire subband edges due to an applied
magnetic field is discussed in Sec.~\ref{spinsplitting}. We provide the theoretical
description of bulk-material and bound-state orbital interactions with a magnetic field 
in Sec.~\ref{theoryBfield}, followed by a detailed analysis of their individual 
contribution to the total $g$ factor of hole-nanowire states in Secs.~\ref{bulkZeeman} 
and \ref{Orbital}. The effects of lower-symmetry corrections and finite $k_{z}$ will be 
considered in Secs.~\ref{cubic} and \ref{finitekz}. We discuss the important interplay
between bulk-material and orbital contributions to the total $g$ factors of hole
nanowire states in Sec.~\ref{interplay}, demonstrating a versatile tunability of the
magnitude and sign of the $g$ factor by simple confinement engineering. We provide
a summary and conclusions in Sec.~\ref{summary}.

\section{Theoretical description of hole nanowire subbands in the absence of a magnetic field}
\label{theory}
\subsection{Luttinger model for bulk semiconductor valence bands}
The electronic structure of bulk valence bands can be described using the Luttinger 
Hamiltonian\cite{luttingerPR1956}
\begin{widetext}
\begin{equation}
H_{L} = -\frac{\hbar^{2}}{2m_{0}} [(\gamma_{1}+5\gamma_{2}/2 ) \hat{k}^{2}- 2
\gamma_{2}(\hat{k}_{x}^{2}\hat{J}_{x}^{2}+ \mathrm{cp}) - 4\gamma_{3}(\lbrace
\hat{k}_{x},\hat{k}_{y} \rbrace \lbrace \hat{J}_{x},\hat{J}_{y} \rbrace + \mathrm{cp})]~~,
\label{luttingerH}
\end{equation}
\end{widetext}
where $m_{0}$ is the vacuum electron mass, $\hat{k}_{i}$ are the components of
linear orbital momentum, $\hat{J}_{i}$ are the Cartesian angular-momentum
operators for a particle with spin $\frac{3}{2}$, and $\gamma_{i}$ the Luttinger parameters. In Eq.\ (\ref{luttingerH}), we have used the notation $\lbrace A,B \rbrace=\frac{1}{2}(AB+BA)$. We will choose our quantization axis for total (spin-3/2) angular 
momentum to be along the $z$-direction. In this representation, $\hat{J}_{z}$ will be 
diagonal with eigenvalues $\pm \frac{3}{2}$ and $\pm \frac{1}{2}$. The majority of the
results presented in this paper focus on the properties of subband edge properties at 
$k_{z}=0$ and, hence, we restrict our discussion of the theoretical formalism to this 
case in the following unless otherwise indicated. At the end of the paper, we will 
discuss hole-nanowire dispersions as a function of $k_{z}$, as well as address the
issue~\cite{perssonNANOLETT2004,zhangEPJB2006} of subband edges occurring
at finite $k_{z}$.

To provide clearer interpretations of our results, we will initially use the Luttinger
Hamiltonian in the spherical approximation~\cite{baldereschiPRB1973} where band
warping terms due to cubic-symmetry corrections~\cite{baldereschiPRB1974} are
neglected. Formally, the spherical approximation consists of setting $\gamma_{2}\to
\gamma_{s}$ and $\gamma_{3}\to\gamma_{s}$ with $\gamma_{s}=(2\gamma_{2}+3
\gamma_{3})/5$. At the end of the paper, we will explicitly discuss the effects of cubic
corrections. In the spherical approximation (and for $k_{z}=0$), the bulk Hamiltonian 
(\ref{luttingerH}) simplifies to
\begin{widetext}
\begin{equation}
H_{L}^{s} = -\frac{\hbar^{2}}{2m_{0}}[\gamma_{1}\hat{k}_{\perp}^{2}\cdot \openone_{4
\times 4} +\gamma_{s}\hat{k}_{\perp}^{2}(\hat{J}^{2}_{z}-\frac{5}{4}\cdot \openone_{4
\times 4}) -  \gamma_{s}(\hat{k}^{2}_{-}\hat{J}_{+}^{2}+\hat{k}_{+}^{2}\hat{J}_{-}^{2})]
~~,
\label{luttingerHspher}
\end{equation}
\end{widetext}
where $\hat{k}_{\pm}=\hat{k}_{x}\pm i\hat{k}_{y}$, $\hat{k}_{\perp}^{2}=\hat{k}_{x}^{2}
+\hat{k}_{y}^{2}$, and $\hat{J}_{\pm}=\left( \hat{J}_{x}\pm i\hat{J}_{y} \right)/\sqrt{2}$.
The parameter $\gamma_{s}$ is a measure of the strength of the SO coupling that is
responsible for lifting the four-fold degeneracy for $k\neq 0$. This arises through the
second term in Eq.\ (\ref{luttingerHspher}), which describes the {\em energy splitting}
between HH and LH bands. The third term is responsible for HH-LH {\em mixing}
through the operators $\hat{J}_{\pm}$, which couple HH and LH amplitudes of the
hole wave function. Thus, for $k_{\perp}\neq 0$ and finite $\gamma_{s}$, the bulk
valence bands can no longer be identified as being of purely HH or LH character in
the sense of having definite spin projections $J_{z}=\pm \frac{3}{2}$ or $J_{z}=\pm
\frac{1}{2}$~\cite{csontosPRB2007}. In the following  discussions, we will restrict the
use of the HH and LH labels solely for states that are pure HH or LH states in this
sense. In the presence of a two-dimensional confining potential, such as in a quantum
wire along the $z$ direction, even states at the band edges have a finite $k_{\perp}$
due to quantum confinement and thus are HH-LH mixtures~\cite{bastardrev,
csontosPRB2007}. 

\subsection{Description of hole states in nanowires}
We will now provide a theoretical description for subbands in hole nanowires confined by potentials with different strengths and confining potentials. We will start off by studying
cylindrical hole nanowires confined by an infinite, hard-wall potential of radius $R$, which is 
particularly interesting due to its symmetry. We note that such wires can be readily fabricated
by self-assembly~\cite{samuelsonNANOTECH2006}.

In the spherical approximation, the sum of the angular momenta $\hat{\mathbf J}$ of 
the band-edge Bloch functions and $\hat{{\mathbf L}}$ of the envelope functions is a 
constant of the motion.\cite{baldereschiPRB1973,baldereschiPRB1974} This 
conserved operator $\hat{{\mathbf F}}=\hat{{\mathbf J}}+\hat{{\mathbf L}}$, which is 
similar to a total angular momentum, was first used by Baldereschi and Lipari~\cite
{baldereschiPRB1973,baldereschiPRB1974} to simplify the acceptor-state problem 
utilizing an analogy between $\hat{{\mathbf L}}$ and $\hat{{\mathbf J}}$ and the {\em 
L-S} coupling scheme in atomic physics. Sercel and Vahala~\cite{sercelPRB1990,
sercelAPL1990} extended this approach to study spherical quantum dots and
cylindrical wires. Their formalism will be the starting point for our own investigations of 
the properties of hole nanowires.

We take the quantum-wire axis to be parallel to the spin-3/2 quantization axis (i.e.,
the $z$ direction). The wire Hamiltonian $H_{{\mathrm L}}^{s}+V_{\text{ch}}(r)$
[including the cylindrical hard-wall potential: $V_{\text{ch}}(r)=0$ for $r<R$ and
$V_{\text{ch}}(r)=\infty$ elsewhere] has common eigenstates with $\hat{F}_{z}=
\hat{J}_{z}+\hat{L}_{z}$. Furthermore, it is a property of the Luttinger Hamiltonian that, 
at $k_{z}=0$, each block labeled by a fixed quantum number $F_{z}$ further 
decouples into two $2\times 2$ blocks according to
\begin{eqnarray}
(H_{{\mathrm L}}^{s})_{F_{z}} = \left [ 
\begin{array}{cc}
H^{+}_{F_{z}} & 0 \\
0 & H^{-}_{F_{z}}
\end{array}
\right ]~~,
\label{blockdiagonal}
\end{eqnarray}
where the $H^{\sigma}_{F_{z}}$ are $2\times 2$ matrices acting in the Hilbert
subspaces spanned by states with spin projections $J_{z}=\left ( \frac{3}{2},-\frac{1}{2}\right )$ 
(for $\sigma=+$) and $J_{z}=\left ( \frac{1}{2},-\frac{3}{2}\right )$ (for $\sigma=-$),
respectively~\cite{sercelPRB1990}.

To determine bulk-hole states at zero magnetic field ($B_{z}=0$) that can be used to
construct cylindrical-wire subband states, we use polar coordinates $(r,\phi)$ and the
wave function ansatz\cite{sercelPRB1990, csontosPRB2007}
\begin{eqnarray}
\psi(r,\phi)=e^{iF_{z}\phi}\left ( 
\begin{array}{c}
a_{F_{z}}^{(\text{ch})}J_{F_{z}-3/2}(k_{\perp}r)e^{-3i\phi/2} \\
b_{F_{z}}^{(\text{ch})}J_{F_{z}+1/2}(k_{\perp}r)e^{i\phi/2} \\
c_{F_{z}}^{(\text{ch})}J_{F_{z}-1/2}(k_{\perp}r)e^{-i\phi/2} \\
d_{F_{z}}^{(\text{ch})}J_{F_{z}+3/2}(k_{\perp}r)e^{3i\phi/2}
\end{array}
\right )~~,
\label{ansatz}
\end{eqnarray}
where $J_{n}(k_{\perp}r)$ is an integer Bessel function and $a^{(\text{ch})}_{F_{z}}
\dots d^{(\text{ch})}_{F_{z}}$ are constants. Diagonalization of $(H_{L}^{s})_{F_{z}}$ 
in this representation yields the following bulk-hole eigenstates
\begin{eqnarray}
\psi_{+F_{z}\nu}(r,\phi) =  e^{iF_{z}\phi} \left (
\begin{array}{c}
a_{\nu}^{(\text{ch})}J_{F_{z}-3/2}(k_{\perp}^{\nu}r)e^{-3i\phi/2} \\
b_{\nu}^{(\text{ch})}J_{F_{z}+1/2}(k_{\perp}^{\nu}r)e^{i\phi/2} \\
0 \\
0
\end{array} \right )~~,
\label{bulkstates_+}
\end{eqnarray}
and 
\begin{eqnarray}
\psi_{-F_{z}\nu}(r,\phi) = e^{iF_{z}\phi} \left (
\begin{array}{c}
0 \\
0 \\
b_{\nu}^{(\text{ch})}J_{F_{z}-1/2}(k_{\perp}^{\nu}r)e^{-i\phi/2} \\
a_{\nu}^{(\text{ch})}J_{F_{z}+3/2}(k_{\perp}^{\nu}r)e^{3i\phi/2}
\end{array} \right )~~.
\label{bulkstates_-}
\end{eqnarray}
In Eqs.\ (\ref{bulkstates_+}) and (\ref{bulkstates_-}) the subscript index $\nu=\pm$
labels eigenstates {\em within} the $2\times 2$ subspaces spanned by HH and LH
states with spin projections $J_{z}=\pm \frac{3}{2}$ and $J_{z}=\mp \frac{1}{2}$, 
respectively. We emphasize that, for $k_{\perp}\neq 0$, hole bands are no longer 
of a purely HH or LH type, hence we use the generic label $\nu=\pm$. 

The coefficients $a_{\nu}^{(\text{ch})},b_{\nu}^{(\text{ch})}$, which do not depend on
$F_{z}$ (nor on $\sigma$ at $B_{z}=0$), are given by~\cite{sercelPRB1990}
\begin{equation}
a_{\nu}^{(\text{ch})}=-\frac{1}{\sqrt{3}}~~;~~~~b_{\nu}^{(\text{ch})}=\sqrt{3}~~.
\end{equation}
The corresponding eigenenergies are
\begin{equation}
E_{\nu}(k_{\perp})=-\frac{\hbar^{2}k_{\perp}^{2}}{2m_{0}}(\gamma_{1}-2\nu \gamma_{s})~~.
\end{equation}

A generalization of these results for finite $z$ component of the magnetic field was
presented in Ref.~~\onlinecite{csontosPRB2007}. There it was pointed out that an
interesting crossover occurs with increasing $k_{\perp}$ for the effective $g$ factor
of bulk hole valence bands in the presence of a finite $B_{z}$  (see Fig.\ 2 of
Ref.~~\onlinecite{csontosPRB2007}). For $k_{\perp}=0$, the spin splitting of the
bulk-hole bands is characterized (in absolute terms) by the $g$ factors $g=6\kappa$
and $g=2\kappa$, which follows from the form of the bulk-hole Zeeman Hamiltonian
that will be introduced in Eq.\ (\ref{zeemanH}). However, at large $k_{\perp}$, spin
splitting turns out to be characterized by $g=0$ and $g=4\kappa$. This situation is
reminiscent of the Zeeman effect in two-dimensional hole systems subject to an
in-plane magnetic field~\cite{winklerbook}, a result that will also be significant for our
later discussion of the spin splitting of a particular class of cylindrical hole-nanowire
subband edges.

Quantum wire eigenstates with subband index $\alpha$ are formed by superimposing
bulk-hole eigenstates according to
\begin{equation}\label{wirestates_cylinder}
\Psi_{\sigma F_{z}}^{(\alpha,\text{ch})}(r,\phi)=c_{\sigma F_{z}+}^{\alpha}
\psi_{\sigma F_{z}+}+c_{\sigma F_{z}-}^{\alpha} \psi_{\sigma F_{z}-} \quad .
\end{equation}
The subband eigenenergies $E^{(\alpha, \text{ch})}_{\sigma F_{z}}$ and expansion
coefficients $c_{\sigma F_{z} \nu}^{\alpha}$ are obtained by imposing the hard-wall
boundary condition $\Psi_{\sigma F_{z}}^{(\alpha,\text{ch})}(R,\phi)=0$ and solving
the secular equations
\begin{widetext}
\begin{eqnarray}
\label{sec+}
a_{+}^{(\text{ch})}b_{-}^{(\text{ch})}J_{F_{z}-3/2}(k_{\perp}^{+}R)J_{F_{z}+1/2}
(k_{\perp}^{-}R)-a_{-}^{(\text{ch})}b_{+}^{(\text{ch})}J_{F_{z}-3/2}(k_{\perp}^{-}R)
J_{F_{z}+1/2}(k_{\perp}^{+}R) & = & 0 \\
\label{sec-}
b_{+}^{(\text{ch})}a_{-}^{(\text{ch})}J_{F_{z}-1/2}(k_{\perp}^{+}R)J_{F_{z}+3/2}
(k_{\perp}^{-}R)-b_{-}^{(\text{ch})}a_{+}^{(\text{ch})}J_{F_{z}-1/2}(k_{\perp}^{-}R)
J_{F_{z}+3/2}(k_{\perp}^{+}R) & = & 0~~.
\end{eqnarray}
\end{widetext}
Again, a generalization for the case of finite magnetic field was presented earlier in 
Ref.~~\onlinecite{csontosPRB2007}. Hole nanowire subband edges can thus be 
obtained in a semi-analytical fashion for a cylindrical hard-wall confining potential.

We have also investigated hole nanowires defined by confining potentials with other symmetry 
and strength, in order to gauge the universality of our findings. In particular, we have focused
on two additional cases, hard-wall confined hole wires with {\em square} (and rectangular) 
crossection, and cylindrical hole nanowires confined by a {\em harmonic} potential.

The first case corresponds to a confining potential that is given by
\begin{eqnarray}
V_{\text{rh}}(x,y) & = & 0~~~~\mathrm{for}~~0\leq x < W_{x}~~\mathrm{and}~~0\leq
y < W_{y}\nonumber \\
V_{\text{rh}}(x,y) & = & \infty~~~~\mathrm{elsewhere}~~~~, \nonumber
\end{eqnarray}
where $W_{x,y}$ are the lateral wire dimensions and $W_{x}=A\cdot W_{y}$. We will 
first consider a square-crossection case with $A=1$, deferring a discussion of the
dependence on aspect ratio $A$ to the end of our paper. Using a complete
two-dimensional square-well basis set, the wave function of each individual wire
subband edge with index $\alpha$ is expanded according to
\begin{eqnarray}
\Psi^{(\alpha,\text{rh})}(x,y)=\sum_{m,n} 
\left (
\begin{array}{c}
a^{(\alpha,\text{rh})}_{mn} \\
b^{(\alpha,\text{rh})}_{mn} \\
c^{(\alpha,\text{rh})}_{mn} \\
d^{(\alpha,\text{rh})}_{mn}
\end{array}
\right ) \frac{2 \sin \frac{m\pi x}{W_{x}} \sin \frac{n\pi y}{W_{y}}}{\sqrt{W_{x}W_{y}}}~~,
\label{wfexpsquare}
\end{eqnarray}
from which the hole wire eigenenergies $E^{(\alpha,\text{rh})}$ and expansion 
coefficients in Eq. (\ref{wfexpsquare}) are obtained by numerical diagonalization of
$H_{L}^{s}+V_{\text{rh}}(x,y) $ within this basis. 

Similar calculations are performed for cylindrical wires confined by a harmonic 
potential
\begin{equation}
V_{\text{cs}}(x,y) = \frac{(\gamma_1+ 5\gamma_2/2) \omega^2}{2}(x^2+y^2)~~,
\end{equation}
where $\omega$ is a measure for the softness of the harmonic potential. The
eigenstates of these wires are expanded in a complete basis set using Hermite 
polynomials according to
\begin{widetext}
\begin{equation}
  \Psi^{(\alpha,\text{cs})}(x,y) = \sum_{m,n} \left (
\begin{array}{c}
a^{(\alpha,\text{cs})}_{mn} \\
b^{(\alpha,\text{cs})}_{mn} \\
c^{(\alpha,\text{cs})}_{mn} \\
d^{(\alpha,\text{cs})}_{mn}
\end{array}
\right ) \sqrt{\frac{\omega}{\pi \; m! \; n! \; 2^{m+n}}}H_m({\sqrt{\omega}x})
H_n({\sqrt{\omega}y})e^{-\omega(x^2+y^2)/2}~~~.
\label{wfexpharmonic}
\end{equation}
\end{widetext}
Numerical diagonalization of $ H_{{\mathrm L}}^{s}+V_{\text{cs}}(x,y) $ within this
basis yields the corresponding eigenenergies $E^{(\alpha,\text{cs})}$ and expansion
coefficients in Eq.~(\ref{wfexpharmonic}).

\section{Valence band mixing in hole nanowires}
\label{bandmixing}
\subsection{Cylindrical hole wires with hard-wall confinement}

We will start by analyzing the hole subband edge states in cylindrical hard-wall wires, 
which are of the form given in Eq.\ (\ref{wirestates_cylinder}). To begin with, we turn
our attention to the detailed form of the secular equations (\ref{sec+}) and (\ref{sec-}). 
A special case arises for subbands with $F_{z}=\sigma \frac{1}{2}$. For these values 
of $F_{z}$, the spinor entries in the hole-wire wave functions
[Eq.\ (\ref{wirestates_cylinder})] are proportional to $J_{\pm 1}(k_{\perp}^{\nu}r)$.
Since $J_{-1}(x)=-J_{1}(x)$, the secular equations reduce to $J_{1}(k_{\perp}^{\nu}r)
=0$; a condition that can be satisfied by the individual bulk-hole eigenstates with
$\nu=\pm$ that are part of the superposition in Eq.~(\ref{wirestates_cylinder}). Thus,
for $F_{z}=\sigma \frac{1}{2}$, the {\em wire\/} eigenstates are actually {\em pure bulk
states} with $\nu=+$ or $\nu=-$. Note, however, that these bulk states are still of
mixed HH and LH character for finite (quantized) $k_{\perp}$~\cite{csontosPRB2007},
hence our labeling in terms of generic quantum number $\nu$ and {\em not} in terms
of HH and LH as some authors do, reserving the latter labeling strictly for states that
have definite angular-momentum projections $J_{z}=\pm \frac{3}{2}$ and $J_{z}=\pm
\frac{1}{2}$, respectively. For any other values $F_{z}=-\sigma \frac{1}{2}$ and $\vert
F_{z} \vert > \frac{1}{2}$ the indices of the Bessel functions appearing in Eqs.\
(\ref{sec+}) and (\ref{sec-}) are not equal and the resulting wire states are true
mixtures of the bulk-hole eigenstates $\psi_{\sigma F_{z}+}(r,\phi)$ and $\psi_{\sigma 
F_{z},-}(r,\phi)$, for a given $\sigma$, in order to satisfy the boundary condition set by
the wire confinement. 

\begin{figure}[b]
\includegraphics[width=2.5in]{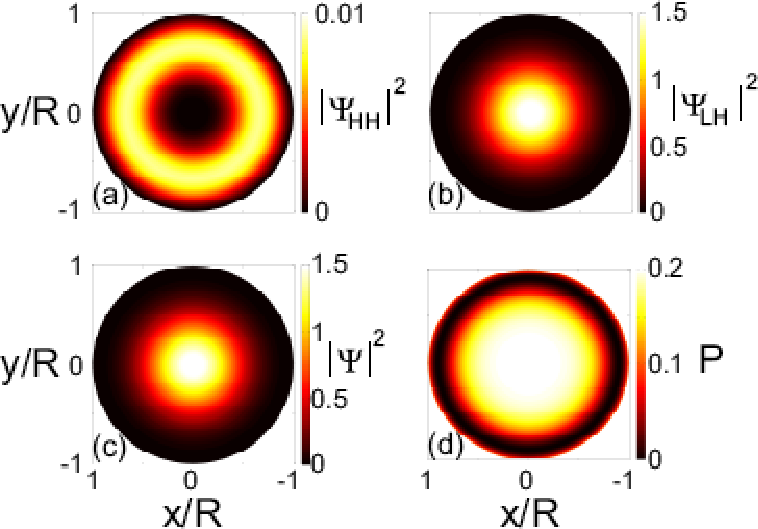} 
\caption{(Color online) Analysis of the highest-in-energy quantum wire subband-edge
state with $\alpha=1$ and $F_{z}=-\frac{1}{2}$ (i.e., $\sigma=+$). Square amplitude of 
(a) the HH component and (b) the LH component of the subband edge wave function.
(c) Total wavefunction amplitude squared. (d) Spatial variation of the hole 
spin-polarization density, $P=\rho_{1}^{2}/\rho_{0}^{2}$. $P=0.2$ corresponds to LH 
and $P=1.8$ to HH character.\label{fig:2}}
\end{figure}

To clarify how strong the HH-LH mixing in fact is, we will analyze the wavefunction of 
hole nanowire subband edges in greater detail. To be specific, we use the Luttinger
parameters characteristic for GaAs~\cite{vurgaftmanJAP2001} in our calculations:
$\gamma_{1}=6.98$, $\gamma_{2}=2.06$, $\gamma_{3}=2.93$, hence $\gamma_{s}
=2.58$. (We remind the reader that the following results are calculated using the
spherical approximation for the Luttinger Hamiltonian.). At $B_{z}=0$, the wire
eigenstates are two-fold degenerate, the degeneracy occuring between states
corresponding to subspaces $\sigma=\pm$ characterized by $\sigma F_{z}$. We will
assume $\sigma=+$ in the following and suppress the index $\sigma$ for brevity.

We illustrate the HH-LH mixing in two different ways. The traditional approach is
to show the squared amplitudes of the HH and LH components in the subband 
eigenfunctions. However, the thus obtained HH or LH density profile depends on the
chosen basis, i.e., the quantization axis of total angular momentum. As an alternative, 
we will also consider a quantity that measures the {\em invariant\/} hole
spin-polarization density, $P=\rho_{1}^{2}/\rho_{0}^{2}$. Here the quantities
$\rho_{1}^{2}$ and $\rho_{0}^{2}$ are scalar invariants derived from the
spin-$\frac{3}{2}$ density matrix~\cite{winklerPRB2004} that are related to the hole
spin-dipole density and total charge density, respectively. A pure HH state has
$P=1.8$ everywhere, whereas a pure LH character yields $P=0.2$ uniformly in space. 
The formulation in terms of scalar invariants is basis-set independent. In principle, it
therefore enables a more rigorous analysis of HH-LH mixing than would be possible in
the traditional approach.

\begin{figure}[t]
\includegraphics[width=1.5in]{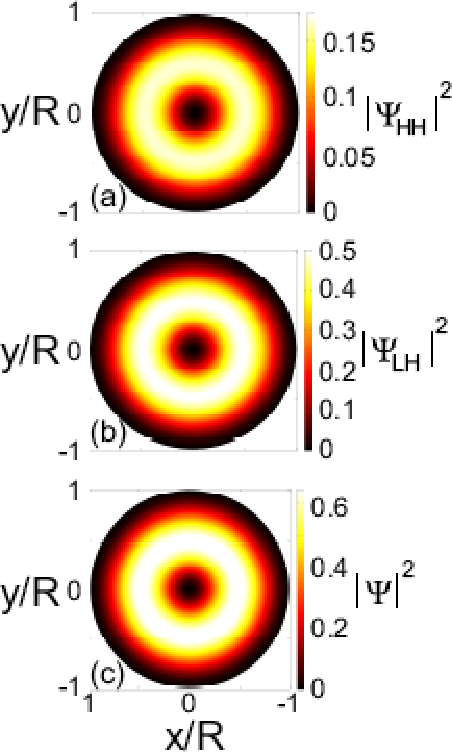} 
\caption{(Color online) Analysis of the quantum wire subband edge with $\alpha=2$ and $F_{z}=\frac{1}{2}$ of a cylindrical, hard-wall confined hole nanowire. Squared amplitude of the (a) HH component, and (b) LH component of the subband edge wave function. (c) Total wavefunction amplitude squared. The hole spin-polarization density,
$P$, is {\em zero} for this subband throughout the entire crossection of the nanowire,
and is thus not shown.\label{fig:3}}
\end{figure}

In Fig.~\ref{fig:2}, results are shown for the topmost subband edge (i.e., the subband
with index $\alpha=1$). The subband-edge state is characterized by $F_{z}=-\frac{1}
{2}$. Figures~\ref{fig:2}(a) and (b) give the cross-sectional profiles of the HH and LH components in the subband-edge wavefunction, illustrating that (i)~the spatial density profile of the two components is very different, and (ii)~their respective overall
magnitudes differ by two orders of magnitude, with the LH component being the
dominant one. These findings are also reflected in the (normalized) {\em hole-spin 
polarization density}, $P$, which is shown in Fig.\ \ref{fig:2}(d). The normalization by
$\rho_{0}^{2}$ (which is a measure of the hole {\em charge} density) ensures that the 
observed spatial variations in the hole-spin polarization density can indeed be 
attributed to hole spin and not charge-density variations. For comparison, the
charge density profile is shown in Fig.~\ref{fig:2}(c).

Inspection of the hole-spin polarization density $P$ shows that the eigenstate is 
predominantly LH-like throughout most of the core of the wire, where $P=0.2$. 
However, toward the edge, i.e., with increasing $r$, the hole-spin polarization {\em 
vanishes}, only to recover to a finite value at the very edge of the cylindrical confining 
potential. This is a manifestation of HH-LH mixing. The predominantly LH character of 
the topmost hole-wire subband in GaAs is in agreement with other theoretical
predictions~\cite{perssonAPL2002,haradaPRB2006,zhuNANOLETT2007}.

\begin{figure}[t]
\includegraphics[width=2.5in]{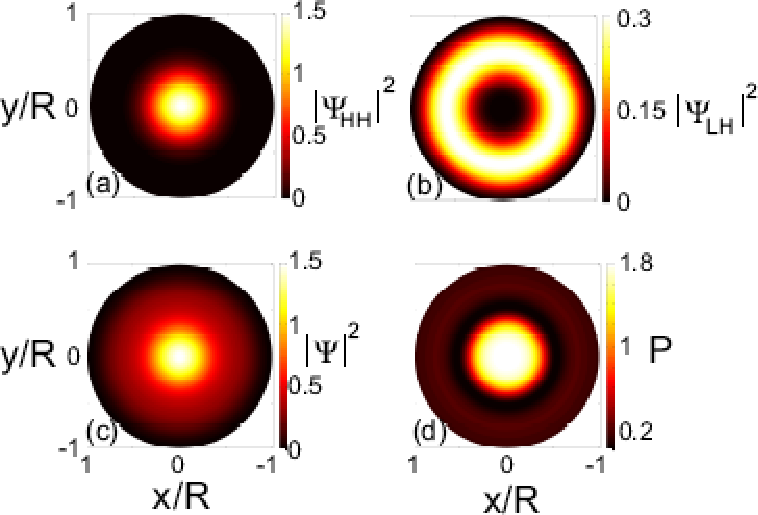} 
\caption{(Color online) Analysis of the quantum wire subband with $\alpha=3$ and $F_{z}=\frac{3}{2}$. Squared amplitude of the (a) HH component, and (b) LH component of the total subband edge wave function. (c) Total wavefunction amplitude squared. (d) Spatial distribution of the hole spin-polarization density, $P=\rho_{1}^{2}/\rho_{0}^{2}$.
\label{fig:4}}
\end{figure}

We now turn our attention to the second-highest subband, with index $\alpha=2$.
This wire subband edge is characterized by $F_{z}=\frac{1}{2}$ and therefore is
actually a bulk-hole eigenstate. (See discussion above.)
Figures~\ref{fig:3}(a)--\ref{fig:3}(c) show the squared amplitudes of the HH and LH
components of the wavefunction, along with the total hole charge density profile. The
hole-spin polarization density turns out to vanish identically across the entire 
crossection of the wire and is therefore not shown. This means that the
second-highest subband edge {\em completely\/} lacks spin polarization. Here the 
difference between the traditional approach toward discussing HH-LH mixing and the 
alternative one based on spin-density-matrix invariants~\cite{winklerPRB2004} is most 
transparent. We will elaborate on this fact more later on.

The third-highest subband edge turns out to be of predominantly HH character, as 
can be seen from Fig.\ \ref{fig:4}. The hole-spin polarization density profile,
Fig.~\ref{fig:4}(d), shows clearly that the wavefunction around the core of the wire is
of purely HH character ($P=1.8$). With increasing $r$, however, the hole-spin 
polarization drops to zero, subsequently recovering values around $P=0.2$, which 
signifies a LH character. The HH-LH mixture variations can also be seen explicitly by 
comparing the magnitudes and spatial profile of the square amplitudes of the HH and 
LH components of the wavefunction, shown in Figs.~\ref{fig:4}(a) and (b),
respectively.

\begin{figure}[b]
\includegraphics[width=3in]{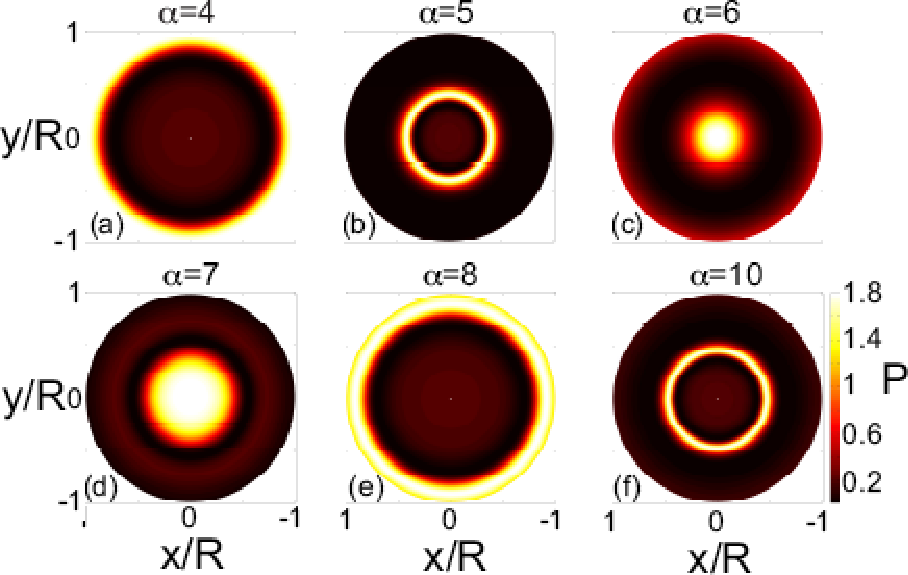} 
\caption{(Color online) Spatial distribution of the hole spin-polarization density, $P=\rho_{1}^{2}/\rho_{0}^{0}$, of subband edge states with (a) $\alpha=4$, $F_{z}=-\frac{3}{2}$ (b) $\alpha=5$, $F_{z}=-\frac{1}{2}$ (c) $\alpha=6$, $F_{z}=\frac{3}{2}$ (d) $\alpha=7$, $F_{z}=\frac{5}{2}$ (e) $\alpha=8$, $F_{z}=-\frac{5}{2}$ (f) $\alpha=10$, $F_{z}=-\frac{3}{2}$. The wire subband with $\alpha=9$ is characterized by $F_{z}=\frac{1}{2}$ and has zero hole-spin polarization throughout the entire crossection of the wire, and is thus not shown.\label{fig:5}}
\end{figure}
For subbands at decreasing energy (i.e., those with higher subband index $\alpha$),
the hole-spin polarization density profiles become increasingly complex. In
Fig.~\ref{fig:5} we show the hole-spin polarization density $P$ for the subband edges 
with $\alpha=4\dots 8$ and $\alpha =10$. The subband edge with $\alpha=9$ has
 $F_{z}=\frac{1}{2}$ similarly to the one with $\alpha=2$ and, hence, is again also
a bulk-hole eigenstate with vanishing polarization across the entire wire crossection.

\begin{figure*}
\includegraphics[width=5.5in]{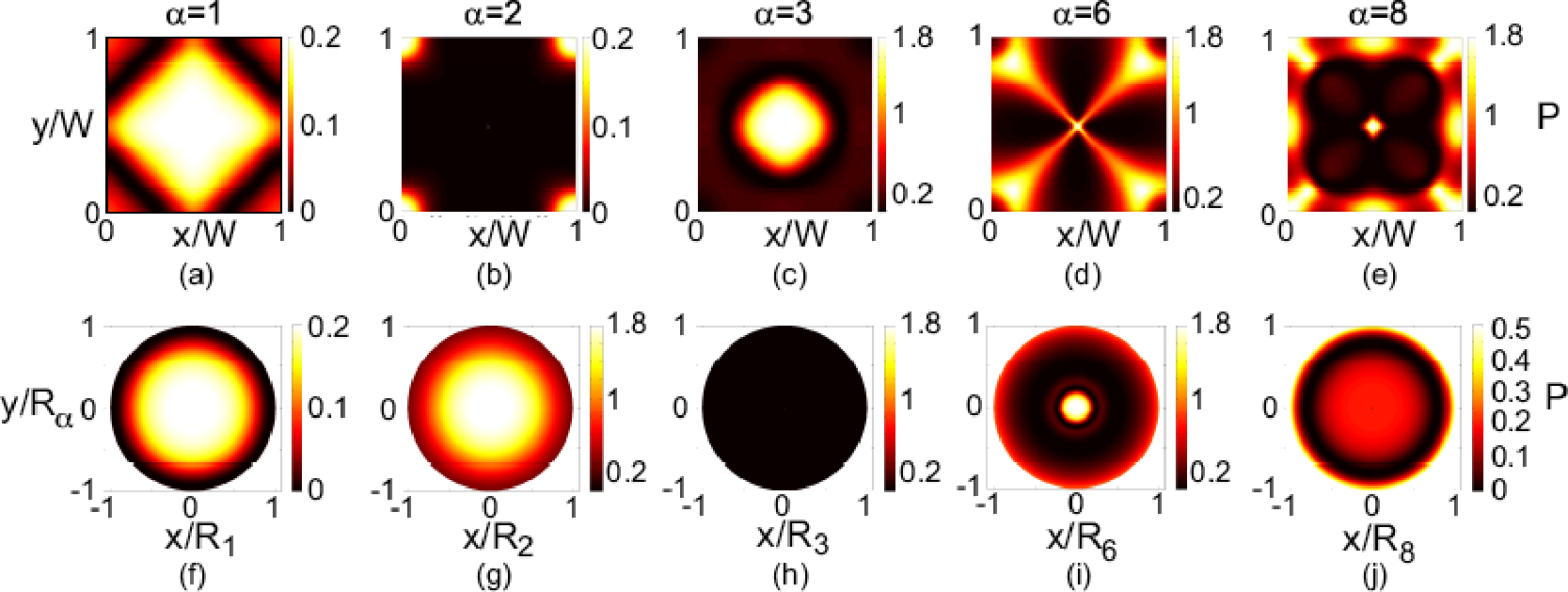}
\caption{(Color online) Spatial distribution of the hole spin-polarization density, $P=\rho_{1}^{2}/\rho_{0}^{0}$ for the levels $\alpha=1,2,3,6,8$, for square-crossection
hard-wall confinement (a)-(e) and cylindrical soft-wall confinement (f)-(j) hole nanowires. $R_{\alpha} = [ 2 E^{(\alpha,\text{cs})}/(\gamma_{1} + 5/2\gamma_{s})
]^{1/2}$. \label{fig:6}}
\end{figure*}

The fourth-highest level ($\alpha=4$) is predominantly LH-like throughout the core of
the wire. The hole-spin polarization drops to zero with increasing $r$, only to recover 
and switch to a HH character, $P=1.8$, at the edge of the wire. A similar behavior is 
observed for the subband edge with $\alpha=8$. The other subband edges show 
equally large variations of the hole-spin polarization as a function of the radial 
coordinate $r$. This has been found to be a general feature for all subband-edge
levels characterized by $F_{z}\neq \frac{1}{2}$ and is a clear manifestation of the strong valence band mixing.

\subsection{Dependence on shape and strength of the wire confinement}
To investigate the universality of our findings, we have performed similar analysis for 
hole nanowires with confining potentials of different shape (or symmetry) and strength. 
In particular, we considered two additional cases, namely hard-wall confined hole
wires with {\em square} crossection, and cylindrical hole nanowires confined by a {\em 
harmonic} potential. We continue using the spherical approximation for the moment,
deferring a discussion about cubic crystal-symmetry effects to later parts of the paper.

The subband edges of wires defined by the two additional types of confinement are
analyzed in the same way as we did previously for the cylindrical, hard-wall confined 
ones. In particular, we are again focusing on the hole-spin polarization density 
profiles to reveal the nature of HH-LH mixing. In Fig.~\ref{fig:6}, we show the
calculated hole-spin polarization density profiles, $P$, for square crossection,
hard-wall (upper row), and cylindrical, soft-wall hole nanowires (bottom row). For brevity, 
we only focus on the subband edges with $\alpha=1,2,3,6,8$ to illustrate our main
findings.

The topmost subband with $\alpha=1$ is seen to be of LH character, i.e., having $P=0.2$, throughout most of the wire crossection. Especially the cylindrical wire with 
harmonic confining potential shows a hole-spin polarization density profile that is very 
similar to the corresponding hard-wall, cylindrical wire. This is not surprising, as both 
confining potentials are axially symmetric. For the square-crossection wire, on the other hand, the lower symmetry of the confining potential is reflected in the 
polarization density profile. Nevertheless, the predominant character of the subband
edge is still LH-like, showing that this feature is indeed very
robust~\cite{haradaPRB2006,zhuNANOLETT2007}.

The second-highest subband edge (with $\alpha=2$) shows quite a different hole-spin polarization profile when comparing the different confining potentials. The square 
crossection nanowire shows a vanishing polarization at the center of the wire, similar 
to the cylindrical hard-wall wire. However, due the reduced symmetry, the polarization 
is finite at the corners of the wire, where the hole-spin polarization is of LH-character. 
For the soft-wall cylindrical wire, the second subband displays a predominantly HH 
character throughout most of the nanowire crossection. However, examining the third-highest 
subband edges with $\alpha=3$, we see that the square crossection,
hard-wall wire state exhibits a HH-character (similar to the cylindrical hard-wall case),
whereas the cylindrical soft-wall nanowire has zero polarization. Thus, a level 
switching in energy between the subband edges with $\alpha=2$ and $\alpha=3$ has 
occured for the soft-wall cylindrical nanowire in comparison with its hard-wall 
cylindrical and square crossection counterparts.

For subband edges with higher $\alpha$, the deviations from axial symmetry for the 
square crossection wire become increasingly more prominent, as is shown explicitly
for the subband edges with $\alpha=6$ and $\alpha=8$. The soft-wall cylindrical wire 
on the other hand displays hole-spin polarization profiles similar to the hard-wall 
cylindrical case. 

Quite generally, our extensive studies of the hole-spin polarization of subband edges
in nanowires show that HH-LH mixing strongly affects the polarization of individual 
subbands, a characteristic that is universal and independent of the symmetry and 
shape of the confining potential. In addition, we find that the topmost subband edge
states with small $\alpha$ have HH-LH characters that are robust against changes in
the confining potentials.

\section{Spin splitting of hole-wire subband edges}
\label{spinsplitting}

\subsection{Luttinger model for bulk holes subject to a magnetic field}
\label{theoryBfield}
We will now focus on the spin splitting of hole nanowire states in the presence of a 
magnetic field applied parallel to the wire axis. The external field couples to both the
spin-3/2 and orbital degrees of freedom. The Zeeman Hamiltonian describing the 
interaction between the magnetic field, $B_{z}$, and the spin degree of freedom is 
given by
\begin{equation} 
H_{Z}=-2\kappa \mu_{B}B_{z}\cdot \hat{J}_{z}~~,
\label{zeemanH}
\end{equation}
where $\mu_{B}$ is the Bohr magneton and $\kappa$ is the bulk hole {\it g}
factor~\cite{winklerbook}. We will neglect the anisotropic Zeeman contribution that
is of higher order in ${\mathbf J}$ because, in typical semiconductors and for the
crystallographic wire direction we consider, it is much smaller than the one given in
Eq.~(\ref{zeemanH}).

In order to include the orbital effects due to a magnetic field applied in the $z$
direction, we replace $\hat{{\mathbf k}}\rightarrow \hat{{\mathbf k}}+e{\mathbf A}$. 
The symmetric gauge ${\mathbf A}=(-\frac{y}{2},\frac{x}{2},0)B_{z}$ will be used,
and we will only consider resulting terms linear in $B_{z}$ because we extract $g$
factors from the small-magnetic-field limit. In the spherical approximation, i.e.,
using the Luttinger Hamiltonian of Eq.~(\ref{luttingerHspher}), the orbital terms
arising from the replacement of canonical wavevector with the kinetic one have the
following form (we use atomic units and the definitions $\hat{L}_{z}=x\hat{k}_{y}-y
\hat{k}_{x}$ and $\hat{x}_{\pm}=x\pm iy$)
\begin{widetext}
\begin{equation}
\label{HOrbspherical}
H_{\mathrm{orb}}^{s} = H_{\mathrm{orb,diag}}^{s}+ H_{\mathrm{orb,mix}}^{s} =
-\left [\gamma_{1}\cdot \openone_{4\times 4} + \gamma_{s}\left ( \hat{J}_{z}^{2} -
\frac{5}{4}\cdot \openone_{4\times 4} \right ) \right ] \hat{L}_{z} \mu_{B}B_{z}
- i\gamma_{s}\left ( \hat{x}_{-}\hat{k}_{-}\hat{J}_{+}^{2}-\hat{x}_{+}\hat{k}_{+}
\hat{J}_{-}^{2}\right ) \mu_{B}B_{z}~~.
\end{equation}
\end{widetext}
The first term is proportional to the orbital angular momentum operator $\hat{L}_{z}$
and is diagonal in spin space. We denote this contribution by $H_{\mathrm{orb,diag}}^{s}$.
The remaining terms are off-diagonal in spin space, coupling states with $J_{z}$ eigenvalue
differing by 2. We denote this contribution by $H_{\mathrm{orb,mix}}^{s}$. These two
contributions will be analyzed in greater detail in our discussion of the orbital contribution
to the total $g$ factor of hole nanowire subband edges.

At zero magnetic field, subbands with a given $\alpha$ and $\sigma F_{z}$ are doubly 
degenerate. A finite $B_{z}$ lifts this degeneracy, giving rise to a spin splitting from
which we extract a $g$ factor according to
\begin{equation} 
g_{\mathrm{tot}}^{\alpha}=\lim_{B_{z}\rightarrow 0} \mathrm{sgn}(\langle \hat F_z\rangle^+_
\alpha)\frac{E_{\alpha}^{+}(B_{z})-E_{\alpha}^{-}(B_{z})}{\mu_{B}B_{z}}  \quad .
\end{equation}
The sign factor introduces the physical definition of assigning sign according to the projection of 
$\langle {\mathbf {\hat F}}\rangle$ along the magnetic field direction. It removes any ambiguity 
relating to the particular choice of symmetry, $\sigma$, labelling the $2\times 2$ blocks of the 
Hamiltonian. Due to the finite energy separation between wire subbands, this definition is 
equivalent to the perturbative (in $B_{z}$) result
\begin{eqnarray}
g_{\mathrm{tot}}^{\alpha} &  = & \mathrm{sgn}(\langle\hat F_z\rangle^+_\alpha)\frac{\langle
H_{\mathrm Z}+H_{\mathrm{orb}}^{s} \rangle_\alpha^+-\langle H_{\mathrm Z}+H_{\mathrm
{orb}}^{s} \rangle_\alpha^-}{\mu_{B}B_{z}}  \quad , \nonumber \\
& \equiv & \mathrm{sgn}(\langle\hat F_z\rangle^+_\alpha)\frac{2\langle H_{\mathrm Z}+
H_{\mathrm{orb}}^{s} \rangle_\alpha^+}{\mu_{B} B_{z}} \quad ,
\label{gpert}
\end{eqnarray}
where the expectation values are taken with respect to the $B_{z}=0$ wire subband
edge eigenstates with $\sigma=+$. 

In the following, we analyze the individual bulk-material and orbital contributions to 
the total $g$ factor $g_{\mathrm{tot}}^{\alpha}=g_{\mathrm{Z}}^{\alpha}+
g_{\mathrm{orb,diag}}^{\alpha}+g_{\mathrm{orb,mix}}^{\alpha}$ that arise from the
linear-in-$B_{z}$ terms $H_{\mathrm{Z}}$, $H_{\mathrm{orb,diag}}^{s}$, and
$H_{\mathrm{orb,mix}}^{s}$, respectively. It is useful to understand each contribution's
separate properties, as their relative importance may vary depending on physical
details. For example, the bulk-material contribution $g_{\mathrm{Z}}^{\alpha}$ will
dominate in diluted magnetic semiconductor wires where $\kappa$ can be enhanced
by several orders of magnitude due to the $p$--$d$ exchange interaction between
hole carriers and magnetic dopant ions~\cite{dietlPRB2001,csontosAPL2008}.

\subsection{Bulk-material contribution to $g$ factors}
\label{bulkZeeman}

We begin by analyzing the contribution to $g^\alpha_{\mathrm{tot}}$ that is due to the
Hamiltonian $H_{\mathrm{Z}}$. It follows from Eqs.~(\ref{gpert}) and (\ref{zeemanH})
that the $g$ factor derived from the bulk-hole Zeeman effect has the form
\begin{equation}
g_{Z}^{\alpha} = -4\kappa~\mathrm{sgn}(\langle\hat F_z\rangle_\alpha^+) \left \langle
\hat{J}_{z} \right \rangle_\alpha^+ =-2 \sqrt{5} \kappa ~ \mathrm{sgn}(\langle\hat F_z
\rangle_\alpha^+)\rho_{10}.
\label{gZshort}
\end{equation}
In the r.h.s\ equality, we have expressed $g_{{\mathrm Z}}^{\alpha}$ in terms of
$\rho_{10}$, which is a dipole-related component of the spin-3/2 density matrix
(defined in Table~III of Ref.~~\onlinecite{winklerPRB2004}).

It is clear from Eq.\ (\ref{gZshort}) that an eigenstate of pure HH character has an
effective $g$ factor of $-6\kappa$, while a pure LH state has a spin splitting
characterized by $g=-2\kappa$. This is certainly the case for bulk-hole states at the
valence-band edges~\cite{winklerbook}. The situation is different for
quantum-confined hole states, e.g., in a wire~\cite{bastardrev,csontosPRB2007}.
As discussed in Sec.~\ref{bandmixing}, strong HH-LH mixing is a universal
characteristic of hole nanowire subband edges, and hole nanowire states of
purely HH or LH character are not expected to be found. In particular, the analysis of 
the squared amplitude of the spin-3/2 dipole moment in Sec.~\ref{bandmixing}
showed that the normalized hole-spin polarization profiles, $P=\rho_{1}^{2}/
\rho_{0}^{2}$ display strong spatial variations as a manifestation of HH-LH mixing.
($\rho_1$ is directly related to $\rho_{10}$; see Ref.~~\onlinecite{winklerPRB2004}
for details.) The question is thus: To what extent does the HH-LH mixing change the
$g$ factors for hole states in a wire as compared with the bulk-material (HH and LH)
values of $-6\kappa$ and $-2\kappa$? 

\begin{figure}[t]
\includegraphics[width=2.5in]{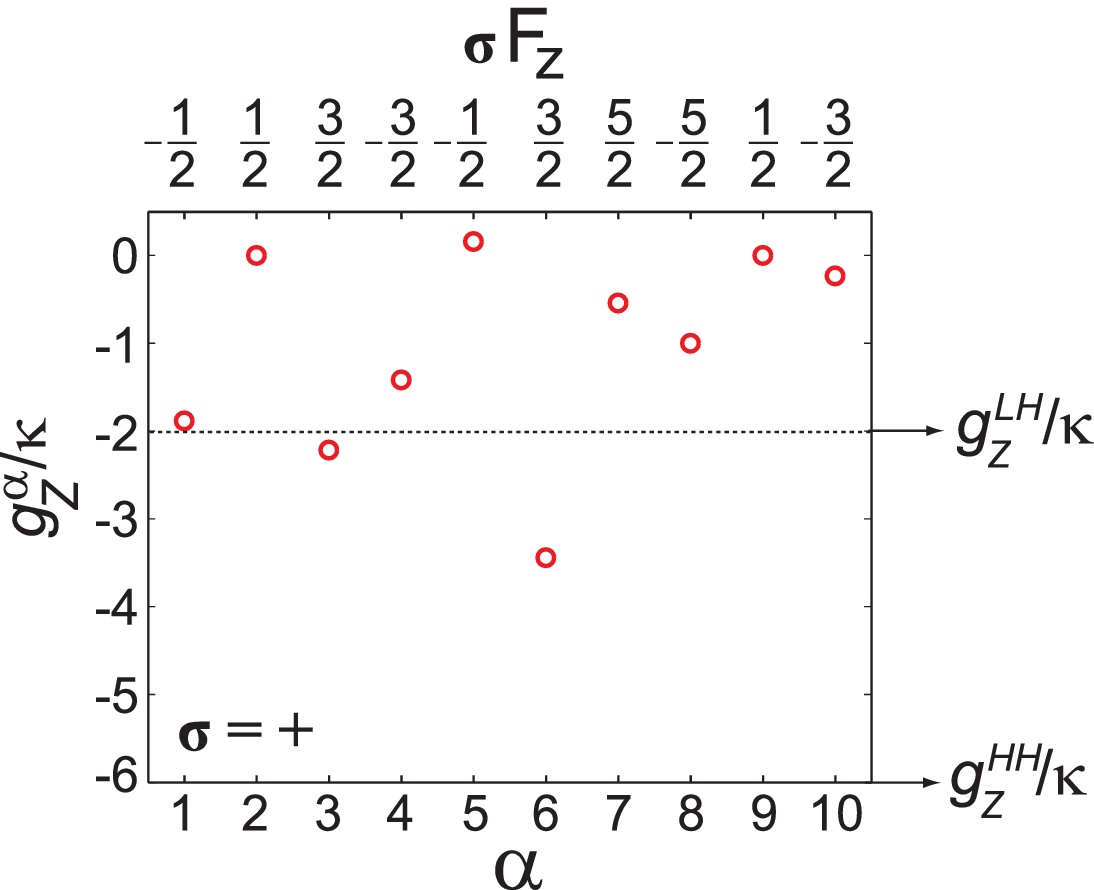} 
\caption{(Color online) Bulk-material contribution to the $g$-factor of the
highest-in-energy subband-edge states, with $\alpha =1\dots 10$, of a hole nanowire
defined by a cylindrical hard-wall confinement. The arrows indicate the values of
$g_{Z}/\kappa =-2$ and $-6$ corresponding to the bulk-hole $g$-factor of pure LH and
HH states, respectively.
\label{fig:1}}
\end{figure}
Figure \ref{fig:1} shows the effective $g$ factors normalized by $\kappa$, of the ten highest subband edges in cylindrical nanowires with hard-wall confinement. The values display strong 
variations as a function of subband index $\alpha$. Furthermore, most subbands have values 
that are strongly suppressed in comparison to the values expected for pure HH and LH states, 
respectively. Inspection of the hole-spin density profiles discussed in Sec.~\ref{bandmixing} and 
comparison with the $g$-factor values shows a strong correlation, establishing a direct 
connection between HH-LH mixing and the observed large $g$-factor variations as a function of 
subband index. For example, the topmost subband edge ($\alpha=1$) was found to be 
predominantly of LH character (see Fig.~\ref{fig:2}). Correspondingly, its $g$ factor is 
found to be close to $-2\kappa$, just as expected from Eq.~(\ref{gZshort}) for a pure 
LH state. 

The second (and ninth) subband edges from the top have vanishing polarizations,
which correlates with a vanishing $g$ factor. Both the second and the ninth state
belong to the class of subband edges with $F_{z}=\frac{1}{2}$ and are thus also a
bulk-hole eigenstate. As discussed in Fig.~2 of Ref.~~\onlinecite{csontosPRB2007},
the spin splitting of bulk hole bands is characterized at high values of $k_{\perp}$ by
values of $g=0$ or $|g|=4\kappa$. The subbands with $\alpha=2$ and $\alpha=9$ are
examples of the former type. We have also found wire-subband edges of the other
type whose spin splitting is characterized by $|g|=4\kappa$, e.g., the one with
$\alpha=16$. 

The dipole moment turns out to be just one of three nontrivial multipole moments that
characterize a spin-3/2 state~\cite{winklerPRB2004}. This is a fundamental difference
with a spin-1/2 system where the dipole moment suffices to uniquely determine the
spin state. When the spin-3/2 polarization (i.e., the dipole moment) vanishes for pure
spin states in two-dimensional~\cite{winklerPRB2004} and one-dimensional~\cite
{csontosPRB2007} hole systems, a substantial {\em octupole\/} moment can exist.
For hole nanowire subband edges, a mathematical relation~\cite{csontosPRB2007}
illustrates this point:
\begin{equation}
\rho_{0}^{2}=2\rho_{2}^{2}=\rho_{1}^{2}+\rho_{3}^{2} \quad .
\label{invrelation}
\end{equation}
Equation~(\ref{invrelation}) links the scalar invariants $\rho_{i}^{2}$ obtained from the
multipole-like expansion of the spin-3/2 density matrix~\cite{winklerPRB2004}. The 
squared monopole, $\rho_{0}^{2}$, and dipole, $\rho_{1}^{2}$, correspond (apart from 
prefactors) to the squared magnitude of the hole charge density and hole-spin 
polarization density, respectively. The quadrupole and octupole are unique to spin-3/2 
systems. The quadrupole is a measure of the HH-LH mixing, while the octupole does 
not have a straightforward physical interpretation. The left equality in
Eq.~(\ref{invrelation}) quantifies the HH-LH mixing that is present at subband edges. 
The right-hand equality implies that the hole-spin polarization, measured by the dipole 
moment $\rho_{1}$, and the octupole, $\rho_{3}$, are complementary. Thus, for the 
subband edges for which the hole-spin polarization vanishes, a maximum octupole 
moment is induced. Similar comparisons can be made for all the wire eigenstates. In 
general, states with mainly HH (LH) character will have $g$ factors close to
$-6\kappa$ ($g=-2\kappa$). States with mixed character, displaying fluctuations in the
hole-spin polarization profiles, have widely varying values for $g$.  

\begin{table}[t]
\caption{Bulk-material contribution to $g$-factors for hole nanowire subband-edges
with index $\alpha=1\dots 10$ for three different types of wire confinement:
(i)~hard-wall cylindrical cross-section, (ii)~hard-wall square cross-section, and
(iii)~soft-wall cylindrical cross-section.}
\begin{tabular}{c|cccccccccc}
$\alpha$& 1 & 2& 3   &  4  & 5 & 6    & 7  & 8  & 9 & 10 \\
\hline
(i) $g_{Z}^{\alpha}$ &  -2.26 & 0.00 & -2.66 & -1.70  & 0.19 & -4.13  & -0.65 
& -1.20  & 0.00 & -0.28 \\
(ii) $g_{Z}^{\alpha}$ &     -2.22 & 0.07 & -3.39  & -1.55  &  0.06 & -2.58   & -0.72 
& -0.59  & 0.08 & -0.17 \\
(iii) $g_{Z}^{\alpha}$ &     -1.84 & -5.57 & 0.00  & -1.34  &  -0.66 & -1.60   & -2.68 
& -0.97  & 0.01 & -0.94 
\end{tabular}
\label{table:1}
\end{table}

\begin{table*}[t]
\caption{Total orbital $g$ factors, $g_{\text{orb}}^{\alpha}$, and the individual
contributions $g_{\text{orb},\text{diag}}^{\alpha}$ and
$g_{\text{orb},\text{mix}}^{\alpha}$ for the ten highest subband edges with $\alpha=1
\dots 10$ in cylindrical hardwall hole nanowires.}
\begin{tabular}{c|cccccccccc}
$\alpha$,$\sigma F_{z}$ & 1,$-\frac{1}{2}$ & 2,$\frac{1}{2}$ & 3,$\frac{3}{2}$   &
4,$-\frac{3}{2}$  & 5,$-\frac{1}{2}$ & 6,$\frac{3}{2}$    & 7,$\frac{5}{2}$  & 8,$-\frac{5}{2}$  &
9,$\frac{1}{2}$ & 10,$-\frac{3}{2}$ \\ \hline
$g_{\text{orb}}^{\alpha}$ &  2.78 & -1.82 & 4.91 & -0.05  & -8.62 & -14.10  & -4.46 
& -3.28  & -1.82 & -18.09 \\
$g_{\text{orb},\text{diag}}^{\alpha}$ &  -0.54 & -1.82 & -8.27 & -12.40  & -10.33 & -5.66  
& -24.10 & -25.04  & -1.82 & -19.63 \\
$g_{\text{orb},\text{mix}}^{\alpha}$ & 3.32 & 0.00 & 13.18  & 12.35  & 1.71 & -8.44   
& 19.64 & 21.76  & 0.00 & 1.54 
\end{tabular}
\label{table:2}
\end{table*}

In Table \ref{table:1} the calculated values of the $g$ factors (with $\kappa=1.2$, the
value~\cite{winklerbook} corresponding to GaAs) for hole wires with
square-crossection hard-wall confinement (mid-row) and cylindrical harmonic potential 
(bottom row) are shown along with the numerical values for the $g$-factors corresponding
to Fig.~\ref{fig:1} for the cylindrical hard-wall nanowire (top row). In general, we observe 
strong similarities between the $g$-factors of the different structures, with respect to 
both magnitude and sign. This is especially clear at low energies, where the axial 
symmetry seems to be retained even in the wire with square crossection. These 
findings do not come as a surprise when one considers the similarity of hole-spin 
polarization density profiles for the low-lying subband edges (shown in
Fig.~\ref{fig:6}). As could be expected, the similarity is stronger for the two types of
cylindrical confinement, while the square-crossection wire states display some
complex hole-spin polarization variations (e.g., for subband edges with $\alpha=6$
and 8), with corresponding discrepancies between the calculated $g$ factor values of
the square crossection wire and the corresponding cylindrical ones. Note also that the
subband edges with vanishing polarization in the cylindrical wires, acquire a finite,
albeit small, polarization for the square crossection one, with a corresponding
non-zero value of the hole $g$ factor. Another interesting observation is the level
switching that occurs between subband edges with $\alpha=2$ and 3 for the cylindrical
wire with parabolic confining potential. This is an example for the sensitivity of
physical quantities (here the wire bound-state energy) with respect to
quantum-confinement-induced HH-LH mixing.

We conclude that, as a universal characteristic of hole nanowires, states are of
mixed HH and LH character even at the subband edges. This is manifested by
large variations in the bulk-material contribution to $g$ factors as a function of 
subband index, concomitant with an overall suppressed magnitude of Zeeman
splitting as compared with free holes.

\subsection{Orbital effects}
\label{Orbital}

We now turn our attention to how the coupling between the holes' orbital degrees of 
freedom and the magnetic field affects the splitting of hole-nanowire subband edges 
in a magnetic field applied parallel to the wire axis. Using a similar perturbation 
approach as in Sec.~\ref{bulkZeeman} , the orbital contribution to the total $g$-factor 
for a subband edge with index $\alpha$ can be written as
\begin{equation}
g_{\mathrm{orb}}^{\alpha}=g_{\mathrm{orb,diag}}^{\alpha}+g_{\mathrm{orb,mix}}^{\alpha}~~,
\label{gpertO}
\end{equation}
where 
\begin{eqnarray}
\label{orbdiag}
g_{\mathrm{orb,diag}}^{\alpha} & = & \frac{-2\left[\gamma_{1}\left \langle \hat{L}_{z}
\openone_{4\times 4} \right \rangle_{\alpha}^+ + \gamma_{s}\left \langle \left ( 
\hat{J}_{z}^{2}-\frac{5}{4} \openone_{4\times 4} \right ) \hat{L}_{z} \right \rangle_{\alpha}^+
\right]}{\mathrm{sgn}(\langle\hat F_z\rangle_{\alpha}^+)}~~, \\
\label{orbmix}
g_{\mathrm{orb,mix}}^{\alpha} & = & \frac{-2\gamma_{s}}{\mathrm{sgn}(\langle\hat F_z
\rangle_{\alpha}^+)}\left \langle i\hat{x}_{-}\hat{k}_{-}\hat{J}_{+}^{2} - i\hat{x}_{+}\hat{k}_{+}
\hat{J}_{-}^{2} \right \rangle_{\alpha}^+. \nonumber \\
\end{eqnarray}

The total orbital $g$-factor contribution $g_{\mathrm{orb}}^{\alpha}$ is shown in
Table~\ref{table:2} (top row) for the ten highest-in-energy subband edges ($\alpha=1
\dots 10$), of a cylindrical hard-wall wire. The orbital $g$ factor displays large 
variations as a function of subband index $\alpha$. Moreover, for some subband 
indices, e.g., $\alpha=6$ and 10, the values are significantly larger than the
corresponding bulk-material contributions discussed in the previous subsection. In
general, we observe that the orbital contribution to the total $g$ factor of hole
subband edges in cylindrical nanowires dominates over the corresponding
bulk-material contributions.

Examining the individual contributions $g_{\mathrm{orb,diag}}^{\alpha}$ and
$g_{\mathrm{orb,mix}}^{\alpha}$, we see that, first of all, the two contributions have 
similar magnitudes. In addition, the two contributions have in general opposite sign, 
with the exception of that for the subband edge with $\alpha=6$. The overall sign of 
the orbital $g_{\mathrm{orb}}^{\alpha}$ is thus determined by the competition between 
these two terms. 

The origin of large fluctuations in the total orbital contribution to the $g$ factor
can be understood by examining the interplay between the diagonal and mixing terms
in greater detail for cylindrical hard-wall hole wires. The operators entering
Eqs.~(\ref{orbdiag}) and (\ref{orbmix}) can be expressed in polar coordinates as
\begin{eqnarray}
\hat{L}_{z}& = & -i\partial_{\phi} \quad , \nonumber \\
\hat{x}_{-}\hat{k}_{-}& = & -i\hat{L}_{-}^{2}(r\partial r+\hat{L}_{z}) \quad , \nonumber \\
\hat{x}_{+}\hat{k}_{+}& = & (\hat{x}_{-}\hat{k}_{-})^{\dagger} \quad , \nonumber \\
\hat{L}_{-} & = & e^{-i\phi}~~.
\end{eqnarray}
Using the wire eigenstates shown in Eq.~(\ref{wirestates_cylinder}) it follows that, in
the limit $B_{z}\rightarrow 0$, the contribution of $H_{\mathrm{orb,diag}}^{s}$ to the
orbital $g$ factor is given by
\begin{widetext}
\begin{equation}
g_{\mathrm{orb,diag}}^{\alpha}= -2~\mathrm{sgn}(\langle\hat F_z\rangle_{\alpha}^+)\left [
\left (F_{z}-\frac{3}{2} \right )\left (\gamma_{1}+\gamma_{s} \right ) \left \langle \frac{3}{2}
\vert \frac{3}{2} \right \rangle_\alpha^+ + \left (F_{z}+\frac{1}{2} \right )\left (\gamma_{1}-
\gamma_{s} \right ) \left \langle -\frac{1}{2}\vert -\frac{1}{2} \right \rangle_\alpha^+ \right ]~~,
\label{gOrbsplitting}
\end{equation}
\end{widetext}
where the angular brackets $\left \langle \frac{3}{2}\vert \frac{3}{2} \right \rangle$ and 
$\left \langle -\frac{1}{2}\vert -\frac{1}{2} \right \rangle$ are squared amplitudes of the 
HH and LH components of an eigenstate $\psi^{(\alpha, \text{ch})}_{+ F_z}$ as described
in Eq.~(\ref{wirestates_cylinder}). Equation~(\ref{gOrbsplitting}) is linear in $F_{z}$, reflecting
the angular momentum that is acquired by the envelope wave function of the hole subband
edge due to the cylindrically symmetric quantum confinement. We also note that the two
terms in Eq.~(\ref{gOrbsplitting}) differ by a factor of $\gamma_{1}\pm \gamma_{s}$. This is
a manifestation of HH-LH {\em splitting}. 

\begin{figure}[t]
\includegraphics[width=2.5in]{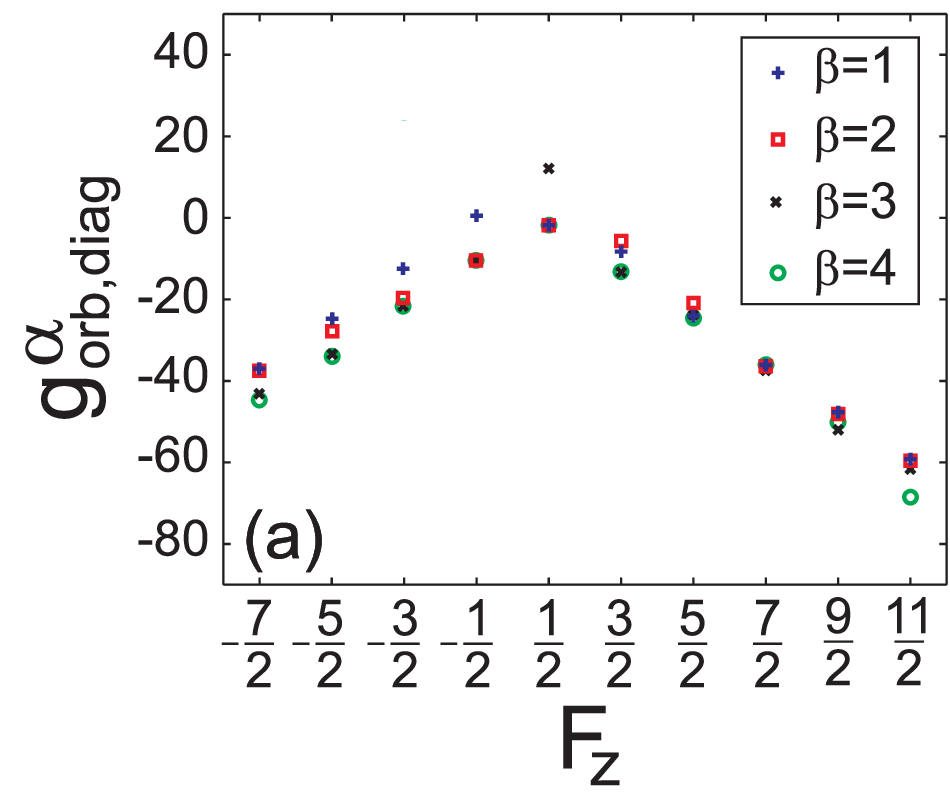} \\[0.2cm]
\includegraphics[width=2.5in]{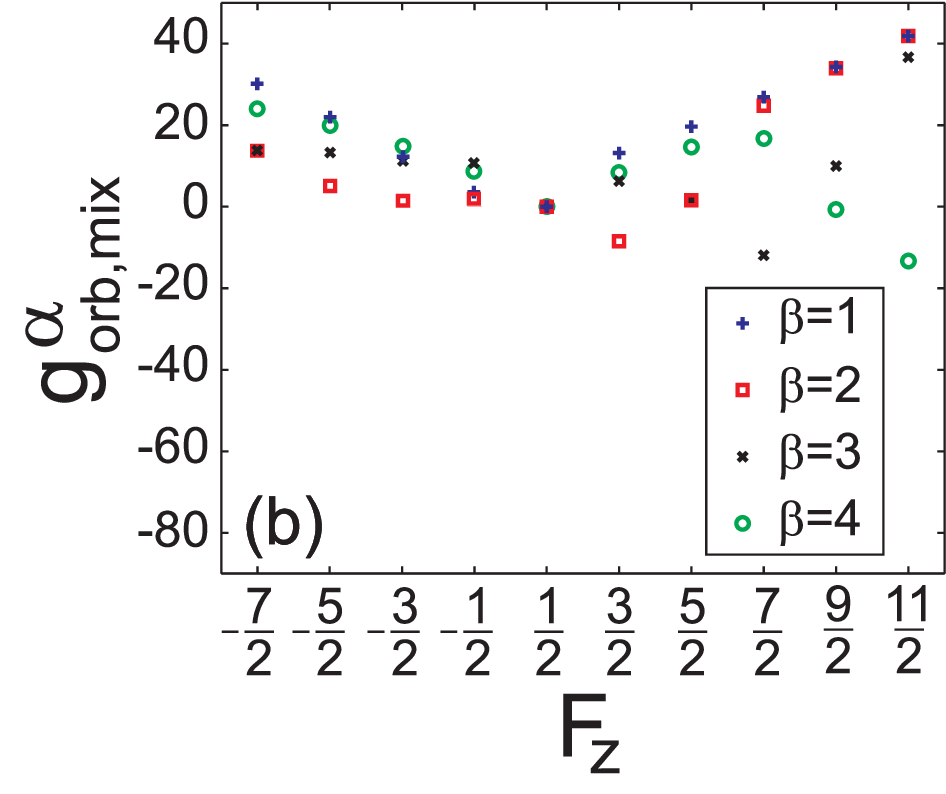}
\caption{(Color online) Values for individual orbital $g$-factor contributions
(a)~$g_{\text{orb},\text{diag}}^{\alpha}$ and (b)~$g_{\text{orb},\text{mix}}^{\alpha}$
for subband edges with $-\frac{7}{2}\leq F_{z}\leq \frac{11}{2}$ (for the $\sigma=+$ states) in
a cylindrical hard-wall hole nanowire. The index $\beta$ labels subband edges with a given
$F_{z}$ in order of decreasing energy. \label{fig:8}}
\end{figure}

In Fig.\ \ref{fig:8}(a), we show values of $g_{\mathrm{orb,diag}}^{\alpha}$ for subband 
edges with different $F_{z}$. For each value of $F_{z}$, the four $g$-factor values correspond
to the four highest-in-energy subband edges with that particular $F_{z}$. 
We label this sequence in energy by the index $\beta$, not to be confused with the 
general subband index $\alpha$. A linear dependence in $F_{z}$ can be clearly seen
from the figure, confirming the presence of the large quantum-confinement-induced 
envelope angular momentum. We also observe a spread in the $g$-factor values for a 
fixed value of $F_{z}$ corresponding to different energies (and subband index
$\alpha$). This is a clear manifestation of HH-LH mixing, which enters through the 
squared amplitudes of the HH and LH components in Eq.~(\ref{gOrbsplitting}). Thus, 
even though the Hamiltonian giving rise to the splitting characterized by
$g_{\mathrm{orb,diag}}^{\alpha}$ is diagonal in spin-space, its contribution to the spin
splitting is affected by the quantum-confinement-induced HH-LH mixing in the 
unperturbed states. The fluctuations in $g_{\mathrm{orb,diag}}^{\alpha}$ as a function 
of subband index $\alpha$ are therefore due to the different subbands having different 
total angular momentum $F_{z}$, as well as different HH-LH character. 

In contrast to $H_{\mathrm{orb,diag}}^{s}$, the Hamiltonian
$H_{\mathrm{orb,mix}}^{s}$ couples hole subband edge wave function components 
with different $J_{z}$. Thus, it gives rise to HH-LH mixing that is in addition to the
HH-LH mixing inherent to the unperturbed wire eigenstates discussed in
Sec.~\ref{bandmixing}. For a cylindrical hard-wall hole wire, this generates a
contribution to the spin splitting given by
\begin{equation}
g_{\mathrm{orb,mix}}^{\alpha} = \frac{-4\sqrt{3}}{\mathrm{sgn}(\langle\hat F_z
\rangle^+_\alpha)}\left \langle \frac{3}{2}\left \vert \hat{L}_{-}^{2} \left ( r\partial_{r} +
\hat{L}_{z}\right  ) \right \vert -\frac{1}{2} \right \rangle_\alpha^+. \label{gOrbmixing}
\end{equation}
Here the notation indicates that the expectation value is between HH and LH 
components of the subband edge state with $\alpha$ and $+F_{z}$, with spin 
projections $J_{z}=3/2$ and $J_{z}=-1/2$, respectively.

In Fig.~\ref{fig:8}(b), we show the $g$-factor contributions
$g_{\mathrm{orb,mix}}^{\alpha}$ for subband edges with $-\frac{7}{2}\leq F_{z} \leq 
\frac{11}{2}$. Again, for each value of $F_{z}$ we show the calculated $g$ factor
values for the four highest-in-energy subband edges with the given $F_{z}$. While 
the characteristics of the mixing term are more complicated, we can discern some 
general trends. For instance, for the lowest-in-energy subband edges, with $\beta=1$, 
a clear linear dependence on $F_{z}$ is seen in the $g$ factors. However, these $g$ 
factors increase monotonically with $|F_{z}|$, in contrast to the splitting term,
$g_{\text{orb},\text{diag}}^{\alpha}$, which is monotonically decreasing with $|F_{z}|$.
In general, the terms $g_{\mathrm{orb,diag}}^{\alpha}$ and
$g_{\mathrm{orb,mix}}^{\alpha}$ have similar magnitudes but opposite sign. However,
for higher energies, this general rule does not apply and a large spread between the
$g$ factors for a given $F_{z}$ at different energies can be observed.

\begin{table}[b]
\caption{Total orbital $g$ factors, $g_{\text{orb}}^{\alpha}$, for the ten highest subband
edges ($\alpha=1\dots 10$) in (i)~cylindrical hard-wall, (ii)~square hard-wall, and
(iii)~cylindrical soft-wall hole nanowires.}
\begin{tabular}{c|cccccccccc}
$\alpha$ & 1 & 2 & 3  &  4  & 5 & 6 & 7 & 8 & 9 & 10 \\
\hline
(i) $g_{\text{orb}}^{\alpha}$ &  2.78 & -1.82 & 4.91 & -0.05  & -8.62 & -14.10  & -4.46 
& -3.28  & -1.82 & -18.09 \\
(ii) $g_{\text{orb}}^{\alpha}$ &  2.74 & -2.21 & 5.88 & -0.69  & -6.26 & -12.12  & -2.39 
& -1.66  & -1.91 & -9.26 \\
(iii) $g_{\text{orb}}^{\alpha}$ & 3.39 & 5.88 & -1.82  & -1.13  & -3.38 & -9.55   & -1.81 
& -4.67  & -1.66 & -6.85 
\end{tabular}
\label{table:4}
\end{table}

A comparison between the orbital contribution to $g$ factors
$g_{\mathrm{orb}}^{\alpha}$ calculated for our three model hole nanowire structures 
is shown in Table \ref{table:4}. Interestingly, with the exceptions of the levels with
$\alpha=2$ and 3, all nanowires have very similar orbital $g$ factors for low-index
levels (small $\alpha$). This is true both for the magnitude and the overall sign of
$g_{\mathrm{orb}}^{\alpha}$. In particular, the square-crossection wire displays similar 
values to the cylindrical ones, despite the lower symmetry of the confining potential. 
Closer inspection of the levels with $\alpha=2$ and 3 also shows that, in fact, the 
similarities extend to these levels as well, if one interchanges the values for $\alpha=
2$ and 3 for the cylindrical wire with harmonic confinement. This is in agreement with 
our previous discussion on the level switching observed in the hole-spin polarization 
of these levels for this wire confinement potential; see Sec.~\ref{bandmixing}. The 
spread between $g_{\mathrm{orb}}^{\alpha}$ factors with given $\alpha$
corresponding to different confining potentials increases, however, with increasing 
subband index $\alpha$, in accordance with the expectation that the more strongly
delocalized wave functions probe the outer edges of the confining potential.

\subsection{Cubic corrections}
\label{cubic}
Next we consider the influence of corrections due to cubic crystal symmetry. These
corrections depend on the orientation of the wire with respect to crystallographic
axes. In the following, we assume that the wire and, hence, the spin-3/2 quantization 
axis, are parallel to the [001] direction. Starting from the full Luttinger Hamiltonian
$H_{L}$ shown in Eq.~(\ref{luttingerH}), we perform the transformation $\hat{k}
\rightarrow \hat{k}+e{\mathbf A}$ in the symmetric gauge. This yields the following
terms linear in $B_z$:
\begin{widetext}
\begin{equation}
\label{HOrbcubic}
H_{\mathrm{orb}} =  -\left\lbrace \left [\gamma_{1}\cdot \openone_{4\times 4} +
\gamma_{2}\left ( \hat{J}_{z}^{2} - \frac{5}{4}\cdot \openone_{4\times 4} \right ) \right ] 
\hat{L}_{z} + \gamma_{2}\left ( y\hat{k}_{x}+x\hat{k}_{y}\right ) \left (\hat{J}_{+}^{2}+
\hat{J}_{-}^{2} \right ) + i\gamma_{3}\left (x\hat{k}_{x}-y\hat{k}_{y}\right )\left (
\hat{J}_{+}^{2} -\hat{J}_{-}^{2} \right ) \right \rbrace \mu_{B}B_{z} .
\end{equation}
\end{widetext}
In contrast to Eq.~(\ref{HOrbspherical}), the orbital Hamiltonian of
Eq.~(\ref{HOrbcubic}) contains terms proportional to $\gamma_{2}$ and
$\gamma_{3}$. The fact that $\gamma_{2}\ne\gamma_{3}$ accounts for bandwarping
effects.  

\begin{figure}[b]
\includegraphics[width=2.5in]{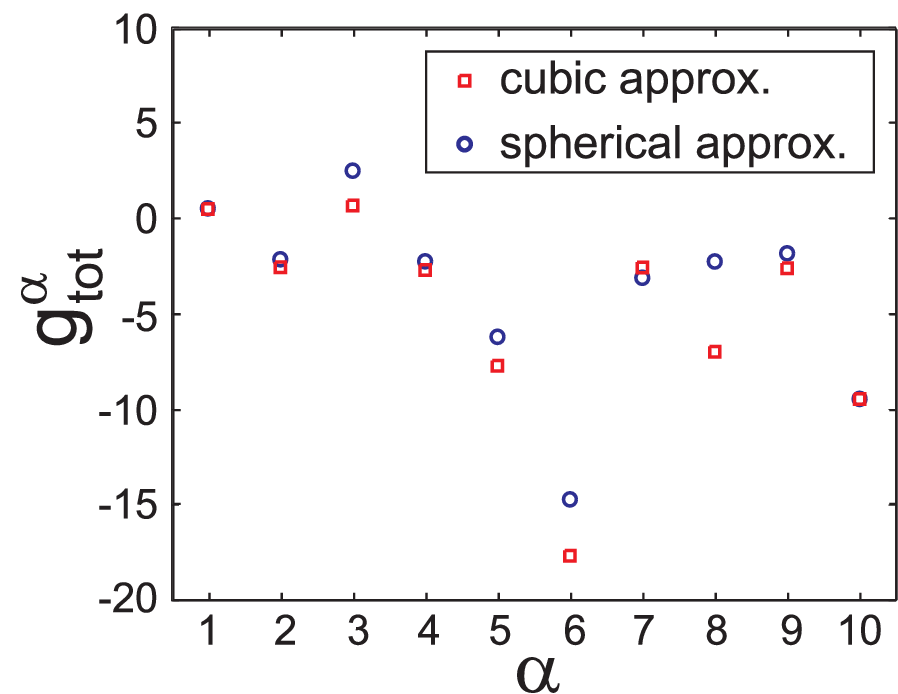}
\caption{Hole $g$ factors $g_{\mathrm{tot}}^{\alpha}$ for the top ten subband edges
($\alpha=1\dots 10$) of a square-crossection hard-wall nanowire. Circles 
correspond to values obtained from diagonalizing $H_{\mathrm{L}}^{s}+
H_{\mathrm{orb}}^{s}+H_{\mathrm{Z}} + V_{\text{rh}}(x,y)$, i.e., using the spherical 
approximation for the Luttinger Hamiltonian~\cite{baldereschiPRB1973}. Squares
denote the values obtained from numerically diagonalizing $H_{\mathrm{L}}+
H_{\mathrm{orb}}+H_{\mathrm{Z}} + V_{\text{rh}}(x,y)$, i.e., using the Luttinger
Hamiltonian reflecting the cubic crystal symmetry. \label{fig:11}}
\end{figure}
In Fig.~\ref{fig:11}, we show the calculated values for $g_{\mathrm{tot}}^{\alpha}$ for 
the hole wire subband edges with $\alpha=1\dots 10$ of a square cross-section
hard-wall nanowire. The open symbols are values obtained from the numerical 
diagonalization of $H_{\mathrm{L}}^{s}+H_{\mathrm{Z}}+H_{\mathrm{orb}}^{s}+
V_{\text{rh}}(x,y)$, i.e., using the spherical approximation for the Luttinger Hamiltonian. 
The filled symbols are obtained by diagonalizing $H_{\mathrm{L}}+H_{\mathrm{Z}}+
H_{\mathrm{orb}}+V_{\text{rh}}(x,y)$, which includes band warping due to the cubic
crystal symmetry. For subband edges with low values of $\alpha$, the $g$ factors
obtained from the two approaches turn out to be very similar. For higher values of
$\alpha$, on the other hand, we observe an increasing spread between the $g$-factor 
values corresponding to the two cases. Properties of wires aligned parallel to
crystallographic directions with lower symmetry than [001] can be expected to be more 
strongly affected by band-warping corrections~\cite{winklerPRL2000}.

\subsection{Interplay of bulk-material and orbital wire-bound-state contributions to
$g$ factors: Tunability of sign and magnitude}
\label{interplay}

We now discuss in greater detail the interplay between the various contributions to
the total nanowire-hole $g$ factor. Lets recall that $g_{\mathrm{tot}}^{\alpha}$ is the
sum of three terms, $g_{\mathrm{tot}}^{\alpha}=g_{\mathrm{Z}}^{\alpha}+
g_{\mathrm{orb,diag}}^{\alpha}+g_{\mathrm{orb,mix}}^{\alpha}$, which are the
bulk-material, diagonal-in-spin-space orbital, and HH-LH mixed orbital contributions,
respectively. In Table \ref{table:3}, we show the calculated $g_{\mathrm{tot}}^{\alpha}$ 
and its components for the ten highest subband edges in hole nanowires with 
square-crossection hard-wall quantum confinement. The values are obtained by 
numerical diagonalization of $H_{\mathrm{L}}+H_{\mathrm{orb}}+H_{\mathrm{Z}}+
V_{\text{rh}}(x,y)$, i.e., using the Luttinger Hamiltonian with cubic (not spherical)
symmetry.

\begin{table*}
\caption{Total $g$ factor $g_{\mathrm{tot}}^{\alpha}=g_{\mathrm{Z}}^{\alpha}+g_{\mathrm{orb,diag}}^{\alpha}+g_{\mathrm{orb,mix}}^{\alpha}$ for the ten
highest-in-energy subband edges ($\alpha=1\dots 10$) of a square-crossection hole 
nanowire with hard-wall confinement. $g_{\mathrm{Z}}^{\alpha}$ is the bulk-material
contribution, which arises from the Zeeman effect of holes in the semiconductor
material. The additional spin splitting due to coupling of the applied magnetic field
to the orbital wire bound state is embodied in $g_{\mathrm{orb,diag}}^{\alpha}$ and
$g_{\mathrm{orb,mix}}^{\alpha}$, which are derived from the first and second terms in
Eq.~(\ref{HOrbcubic}), respectively. Values shown are obtained by numerically 
diagonalizing the cubic Hamiltonian $H_{\mathrm{L}}+H_{\mathrm{orb}}+H_{\mathrm
{Z}}+V_{\text{rh}}(x,y)$ and considering the spin splitting for $B_z\to 0$.}
\begin{tabular}{c|cccccccccc}
$\alpha$ & 1 & 2 & 3 & 4 & 5 & 6 & 7 & 8 & 9 & 10 \\
\hline
$g_{\text{tot}}^{\alpha}$ &  0.50 & -2.55 & 0.67 & -2.69  & -7.71 & -17.64  & -2.56
& -6.97  & -2.61 & -9.45 \\
$g_{Z}^{\alpha}$ &  -2.13 & -0.23 & -4.83 & -1.33  & 0.55 & -1.74  & -1.20
& -0.56  & -0.13 & -0.73 \\
$g_{\text{orb},\text{diag}}^{\alpha}$ &  -0.58 & -2.35 & -3.08 & -12.80  & -10.87 & -11.09  
& -18.89 & -25.99  & -7.16 & -31.20 \\
$g_{\text{orb},\text{mix}}^{\alpha}$ & 3.20 & 0.03 & 8.58  & 11.45  & 2.61 & -4.81 
& 17.52 & 19.58  & 4.68 & 22.47
\end{tabular} 
\label{table:3}
\end{table*}

Table \ref{table:3} illustrates several of our main conclusions. The bulk-material 
contribution shows strong variations in sign and magnitude as a function of subband 
index. This is a manifestation of HH-LH mixing at hole-nanowire subband edges. 
Similar fluctuations occur for the orbital contributions. This is in part due to the fact
that the envelope function of subbands with different $\alpha$ acquire different 
quantum-confinement-induced orbital bound-state angular momenta. In addition, the 
fluctuations also originate from HH-LH mixing. The two orbital contributions
$g_{\mathrm{orb,diag}}^{\alpha}$ and $g_{\mathrm{orb,mix}}^{\alpha}$ have, in 
general, comparable magnitudes but opposite sign. 

\begin{figure}[b]
\includegraphics[width=2in]{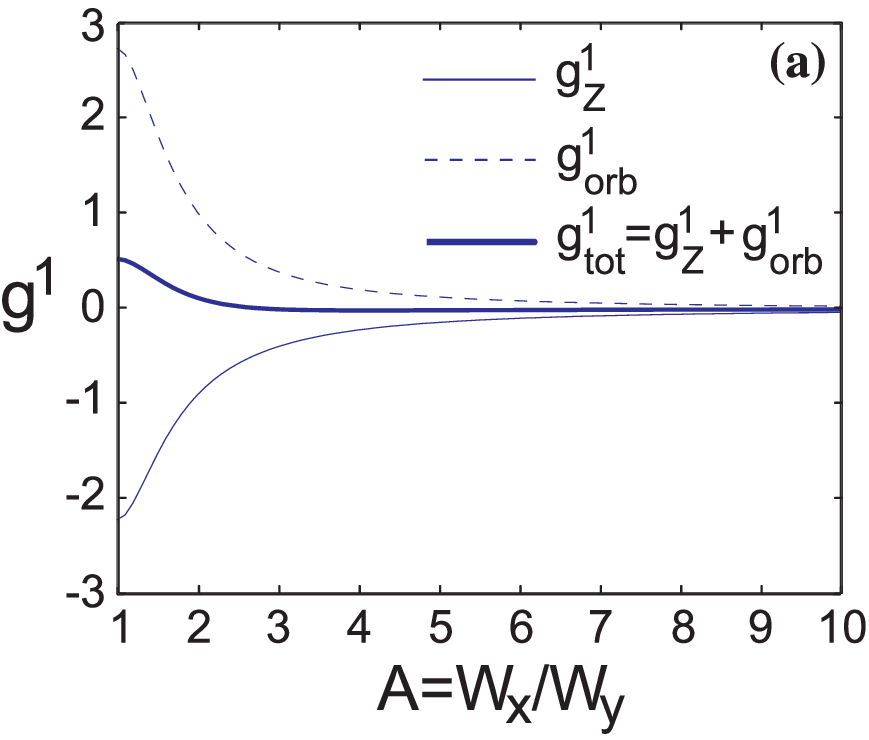} \\
\includegraphics[width=2in]{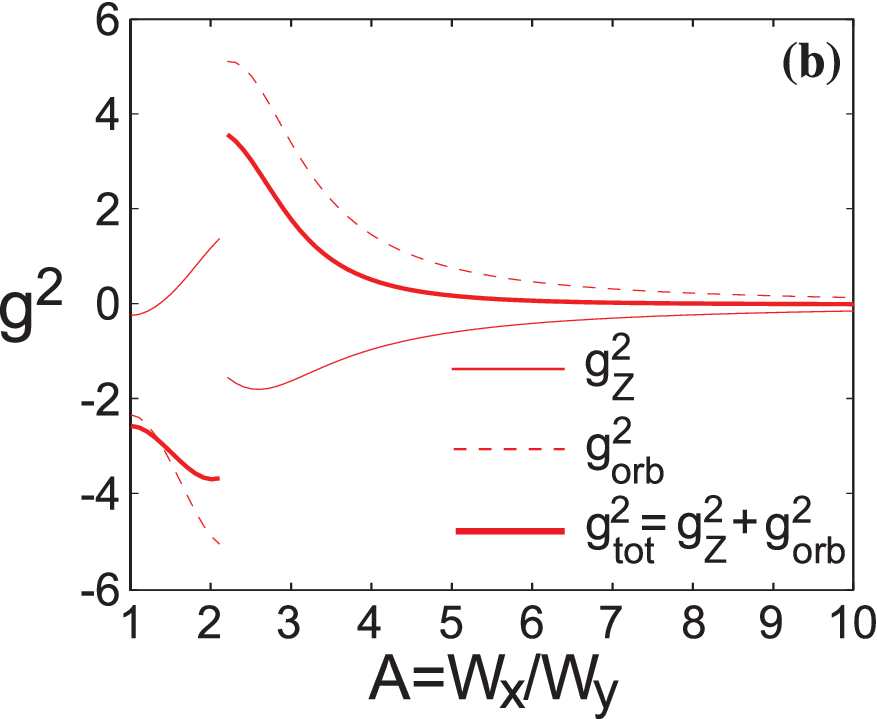} \\
\includegraphics[width=2in]{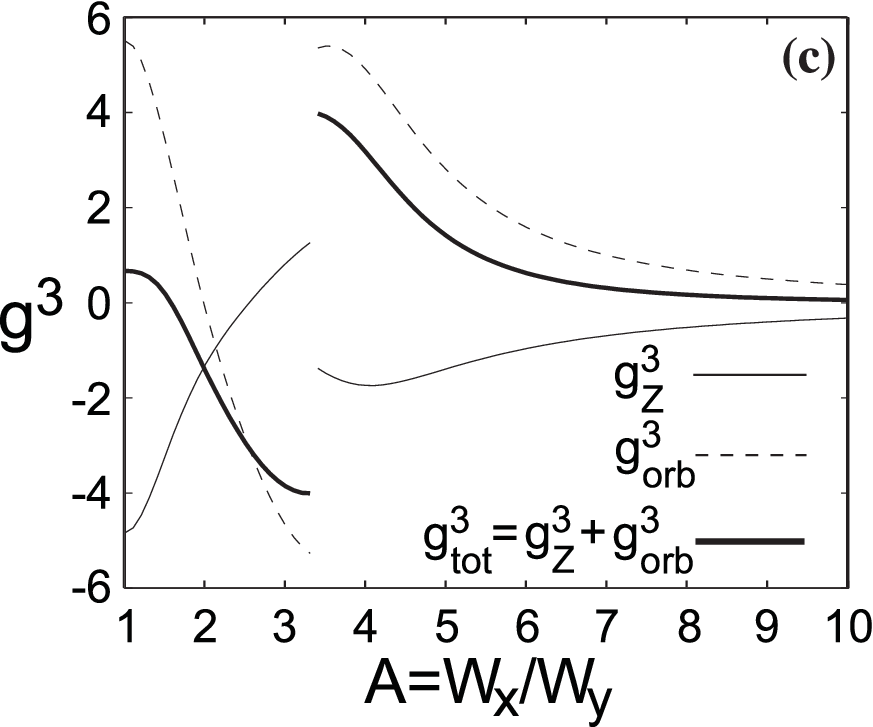}
\caption{Dependence of subband-edge $g$ factors $g_{\mathrm{tot}}^{\alpha}$ on the 
aspect ratio $A$ for a rectangular-crossection hard-wall-confined nanowire (thick solid
curves). We show results for the top three hole-wire subband edges: (a) $\alpha=1$,
(b) $\alpha=2$, and (c) $\alpha=3$. The thin solid and dashed curves are plots of the
corresponding bulk-material and orbital contributions, $g_{\mathrm{Z}}^{\alpha}$ and
$g_{\mathrm{orb}}^{\alpha}$, respectively.\label{fig:10}}
\end{figure}
Most importantly, the total $g$ factor $g_{\mathrm{tot}}^{\alpha}$ depends sensitively
on the relative sign and magnitude of the three contributions. As both the orbital 
contribution to the spin splitting and the detailed form of the HH-LH mixing depend
on the details of the quantum confinement~\cite{kyrchenkoPRB2004}, we expect that
the interplay between the bulk-material and orbital $g$-factor terms can be tuned by
confinement engineering. This would enable the use of nanowires to realize devices
for nanospintronics. In Fig.\ \ref{fig:10} we demonstrate such tunability of hole $g$ factors
in nanowires. The thick curves show $g_{\mathrm{tot}}^{\alpha}$ for the three 
highest-in-energy subband edges of rectangular-crossection hard-wall confined wires 
plotted as a function of the aspect ratio $A=W_{x}/W_{y}$. The bulk-material and
orbital contributions are also shown (by the thin solid and dashed lines, respectively).
A general characteristic that is observed for all three subband edges is that the total $g$
factor approaches zero in the limit of large aspect ratio $A$. This is due to the 
fact that, with increasing $A$, our nanowire system approaches the 2D limit,
essentially mimicking a quantum well subject to an in-plane magnetic field. For such a
system, the topmost hole-quantum-well subband edge is known to have a small
(within the spherical approximation: vanishing) $g$
factor,~\cite{winklerPRL2000, winklerbook} in agreement with our findings. The sudden
sign change observed in $g^2$ and $g^3$ occurs at anticrossing points in the energy
spectrum where $\langle\hat F_z \rangle$ goes through zero, and consequently reverses sign.

The observed dependence of the subband edge $g$ factor values on the aspect ratio 
of the confining potential is quite intriguing. All subband edges display a strong
tunability of the magnitude of the $g$ factor with a change in the aspect ratio. For the
third subband, a variation of $A$ between 1 and 2 gives rise to a change in the {\em sign\/}
of the $g$ factor that is not associated with an anticrossing, thus demonstrating the ability to
gradually tune the spin splitting to zero. The possibility to manipulate both the magnitude and
sign of $g$ factors is a useful ingredient for spintronics applications. We therefore propose
that hole nanowires may be versatile building blocks for nanospintronics due to the
demonstrated tunability that arises from the interplay between bulk-material and orbital spin
splittings in an applied magnetic field. Confinement engineering of nanowires and quantum
point contacts can be readily achieved with present-day technologies, e.g., using side gates
on lithographically defined quantum wires~\cite{wangAPL2000} and quantum point
contacts~\cite{danneauPRL2006, koduvayurPRL2008}, or wrap-gated self-assembled 
nanowires~\cite{ngNANOLETT2004, bryllertEDL2006}. 

\subsection{Hole nanowire dispersions for finite $k_{z}$}
\label{finitekz}

We finish our analysis by discussing the energy dispersion of hole nanowire states
due to motion along the wire direction. For finite $k_{z}$, the decomposition shown in
Eq.~(\ref{blockdiagonal}) is no longer possible. Instead, the full $4\times 4$ Luttinger
Hamiltonian must be considered. We diagonalized numerically the Hamiltonian 
consisting of the sum of $H_{\mathrm{L}}$ shown in Eq.~(\ref{luttingerH}) and
$V_{\text{rh}}(x,y)$ for a square-crossection hard-wall GaAs nanowire oriented along
the [001] direction.

\begin{figure}[t]
\includegraphics[width=2.6in]{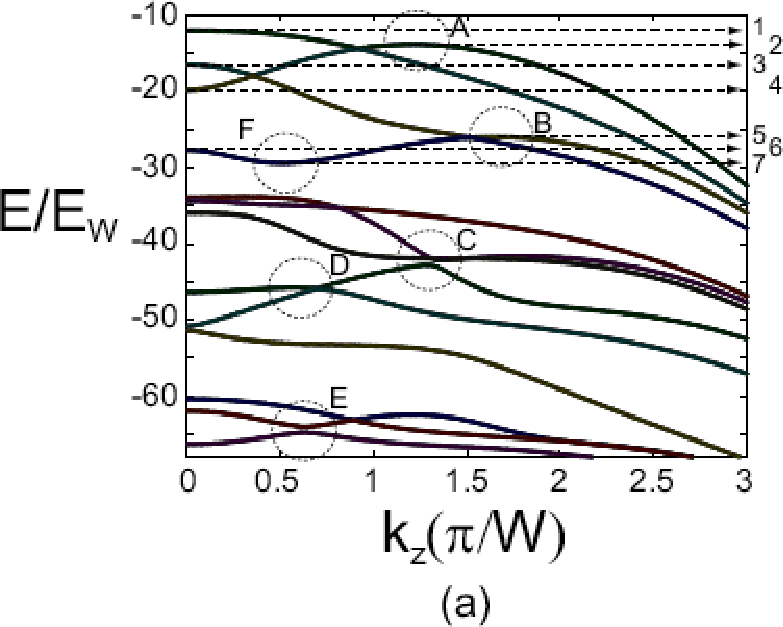}\\[0.2cm]
\includegraphics[width=2.6in]{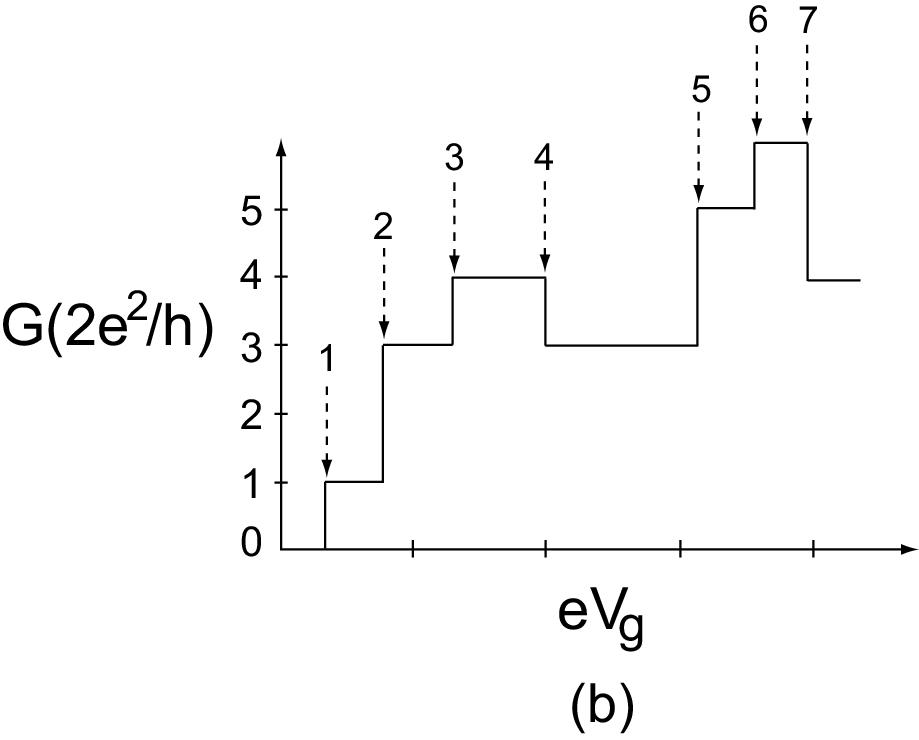}
\caption{(a) Zero-field hole-nanowire subband dispersions for a square crossection 
hard-wall nanowire. The energy has been normalized by $E_{W}=\gamma_{1}
\hbar^{2}/(2m_{0}W_{x}^{2})$. The hole nanowire states were obtained by numerical 
diagonalization of the cubic-symmetry Hamiltonian $H_{\mathrm{L}}+V_{\text{rh}}
(x,y)$. Dashed arrows indicate extremal values of subband dispersions that would be
associated with steps in the two-terminal hole-wire conductance. (b) Schematic
dependence on side-gate voltage for the linear two-terminal conductance through a
hole nanowire or quantum point contact, based on the hole-subband dispersions
shown in (a). Each conductance step corresponds to the Fermi energy reaching a value indicated by the dashed arrows.
\label{fig:12}}
\end{figure}

The highest hole-nanowire subband dispersions are shown in Fig.~\ref{fig:12}. 
Several observations can be made. First, the hole dispersions are strongly
non-parabolic, which is a distinct manifestation of the SO coupling and HH-LH mixing 
in the valence band of typical semiconductors. Second, some subbands display 
electron-like dispersions as well as off-center ($k_{z}\neq 0$) maxima. For example, 
the highlighted points A, B, and D correspond to off-center maxima for the subbands 
with $\alpha=3,4$ and 8, respectively. The occurrence of subband edges away from
the zone center, as well as the strong non-parabolic dispersions, will have direct 
implications for the transport and optical properties of hole nanowires. 

\begin{figure}[t]
\includegraphics[width=2.6in]{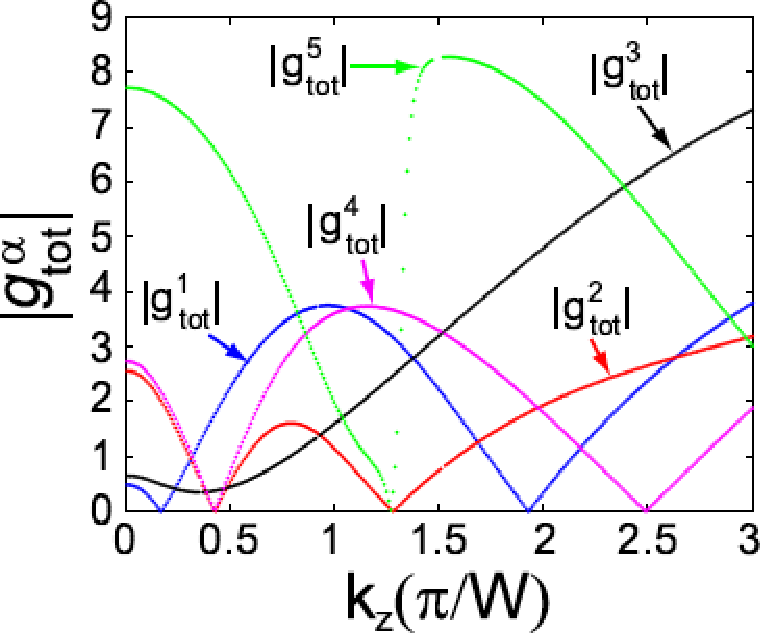}
\caption{(Color online) Absolute values of hole-wire $g$ factors
$g_{\mathrm{tot}}^{\alpha}$ for states at finite wave vector $k_{z}$ for motion parallel
to the wire axis. Results are shown for the subbands with $\alpha=1\dots 5$.
\label{fig:13}}
\end{figure}

One example is the quantized conductance of one-dimensional systems. It is well known that 
transport measurements in the linear-response regime of one-dimensional quantum wires and 
quantum point contacts yield quantized conductance
steps~\cite{vweesPRL1988,wharamJPC1988,wangAPL2000} if the transport is
ballistic. Such conductance steps occur whenever a one-dimensional subband is
opened as a channel for conduction, as it falls below the electrochemical potential
determined by the contacts. Typically, top or side gate are employed to adjust the wire
width and, thus, the quasi-1D subband energies. In the absence of a magnetic field,
the conductance steps occur in units of $2e^{2}/h$.

In Fig.~\ref{fig:12}(b), we show a schematics of the linear-response conductance of a hole 
nanowire with the dispersion shown in Fig.\ \ref{fig:12}(a). The abscissa refers to a gate voltage 
applied to a top or side gate close to the nanowire or quantum point contact device. Application 
of a gate voltage essentially tunes the Fermi level (top gate) or the effective lateral confinement 
(side gate) of the nanowire or quantum point contact device such that different parts of the 
energy spectrum fall within the small energy window around the Fermi level in which conduction 
is possible. In conventional~\cite{vweesPRL1988,wharamJPC1988} zero-field quantized
conductance spectra, the conductance changes monotonically in steps of $2e^{2}/h$ 
as a function of gate voltage. This corresponds to the case of subband edges at
$k_{z}=0$ entering the energy window which is allowed for transport around the Fermi 
level. A hole nanowire can display quite a different conductance spectrum due to its 
strong non-parabolic dispersion. For a hole nanowire with the dispersion show in
Fig.~\ref{fig:12}(a), the conductance can both {\em increase} and {\em decrease} in 
steps of $2e^{2}/h$ and also $2\times(2e^{2}/h)$. This behavior is caused by
off-center subband maxima, examples of which are seen at points A and B in the hole 
dispersion shown in Fig.~\ref{fig:12}(a), as well as switches between hole-like and electron-like 
dispersion, as is the case for point F. A measurement of the transport 
properties in the linear-response regime can thus yield insight to the detailed structure 
of the hole nanowire subband dispersions.

In the presence of a magnetic field, the degeneracy of the hole nanowire subbands is lifted. Similar to the so far discussed spin splitting of the subband edges at $k_{z}=0$, 
the spin splitting of the off-center maxima can also be probed by transport or optical
measurements. For the square-crossection wire with dispersion shown in
Fig.~\ref{fig:12}(a), we have three off-center maxima; at points A, B, and D. 
Calculation of the spin splitting at these three points for a small magnetic field yields
the $g$ factor magnitudes $g^{A}_{\mathrm{tot}}=2.25$, $g^{B}_{\mathrm{tot}}=2.89$, 
and $g^{D}_{\mathrm{tot}}=4.20$, respectively. The full dependence of the $g$ factors of
states in the five highest hole-wire subbands as a function of $k_{z}$ is shown in
Fig.~\ref{fig:13}. Interestingly, the $g$ factor shows a strong dependence on $k_z$;
displaying an oscillatory variation and vanishing intermittently for most subbands.

\section{Discussion and concluding remarks}
\label{summary}

We have calculated Zeeman spitting of states in hole nanowires, subject to a
magnetic field $B_z$ pointing parallel to the wire axis (the $z$ direction) and defined
by three different types of confinement (cylindrical hard-wall, cylindrical soft-harmonic,
and rectangular hard-wall). Our main focus has been to elucidate the properties of
zone-center subband-edge states, although a few results are also given for hole-wire
states having finite wavevector $k_z$ for motion along the wire. We have
disentangled spin-splitting contributions arising from (a)~the bulk-hole Zeeman effect
in the semiconductor material, and (b)~the coupling of the applied magnetic field to
the orbital wire bound states. We termed the contribution (a) the bulk-material
contribution to the $g$ factor and indicated it by the symbol
$g_{\mathrm{Z}}^{\alpha}$, where the index $\alpha$ labels the quasi-1D
hole-wire subbands. The orbital contribution (b) turned out to have two separate
terms. The one indicated by $g_{\mathrm{orb,diag}}^{\alpha}$ reflects the direct
coupling of hole orbital angular momentum component $\hat L_z$ with the magnetic field
and displays HH-LH splitting , and the second one, labelled
$g_{\mathrm{orb,mix}}^{\alpha}$, arises from HH-LH mixing.

Quantum confinement of holes in a wire renders their eigenstates to be coherent
superpositions of HH and LH components. Although these are pure states in the
quantum-statistical sense, the spin polarization of such HH-LH mixed states is not
fixed to be 3/2 as a simple analogy with spin-1/2 systems would suggest. We have
used the recently discussed~\cite{winklerPRB2004} invariant-multipole decomposition
of the spin-3/2 density matrix to universally characterize spin polarization of hole-wire
subband edge states. Within limits, the properties of low index (highest-in-energy)
hole subbands turned out  to be quite robust with respect to changes in wire
cross-sectional shape and confinement strength. Also, many qualitative features are
already captured within models where band-warping due to cubic crystal symmetry
is neglected.

HH-LH mixing is illustrated by radial profiles of the spin-3/2 dipole moment (i.e., spin 
polarization) and turns out to have a profound impact on spin splitting. Variations in 
the spin-polarization profiles between different subband-edge states are reflected in 
fluctuations of the individual $g$-factor contributions, as well as the total
nanowire-hole $g$ factor. It is possible, by addressing different hole-subband
edges, to study nontrivial hole spin-polarization states that have no counterpart in
spin-1/2 systems, such as zero-polarizations states that have a large spin-3/2
octupole component. The bulk-material contribution to the hole-wire $g$ factor could
be investigated separately in wires made from diluted-magnetic semiconductor 
materials, where its magnitude can exceed that of the orbital terms by a few orders of
magnitude.

The spin-splitting properties of individual hole-wire subband edges turn out to be
tunable by changing the aspect ratio in a rectangular-crossection wire geometry.
This is of interest to applications involving electrostatically defined quantum
point contacts or gated self-assembled wires, where the wire shape can be
manipulated \textit{in situ\/}.

Our calculations were performed using Luttinger parameters for GaAs. However,
results obtained within the spherical approximation will apply to other materials
having a similar ratio $\gamma_{\text{s}}/\gamma_1$. The exact quantitative results
will probably be different once cubic corrections are taken into account. However,
as our comparison between fully cubic and approximate spherical models has shown,
qualitative trends for low-index hole-wire subband edges are already captured in the
spherical model. Hence, we believe that our results can be applied to materials other
than GaAs.

The effects of linear-in-wavevector spin-orbit couplings has not been considered.
The relative strength of bulk-inversion-asymmetry (BIA) induced ${\mathbf k}$-linear terms,
as compared with leading-order terms in the Luttinger Hamiltonian, depends strongly
on the material and the direction of motion with respect to crystallographic
axes~\cite{winklerbook}. Given that cubic corrections have not been too important for
low-lying hole-wire subband edges, we expect the same to hold for BIA terms.
Also, while previous studies have shown that, in principle, electric fields applied 
perpendicular to nanowires affect hole $g$ factors~\cite{zhangJPD2007} (essentially
because the HH-LH mixing changes), this effect becomes relevant only at quite large
field magnitudes that are not likely to exist in typical point-contact or nanowire
samples.

\acknowledgments
DC greatfully acknowledges support from the Massey University Research Fund and
also thanks the Division of Solid State Physics/Nanometer Structure Consortium, Lund University, Sweden, for their hospitality and the Swedish Research Council (VR) 
for their financial support during the part of the work that was performed in Sweden.


\begin{thebibliography}{99}
\bibitem{hirumaJAP1995} K.\ Hiruma, M.\ Yazawa, T.\ Katsuyama, K.\ Ogawa, K.\ Haraguchi, and H.\ Kakibayashi, J.\ Appl.\ Phys.\ {\bf 77}, 447 (1995).
\bibitem{gudiksenJPCB2001} M.\ S.\ Gudiksen, J.\ Wang, and C.\ M.\ Lieber, J.\ Phys.\ Chem.\ B {\bf 105}, 4062 (2001).
\bibitem{duanNATURE2001} X.\ Duan, Y.\ Huang, Y.\ Cui, J.\ Wang, and C.\ M.\ Lieber, Nature {\bf 409}, 66, (2001).
\bibitem{TangSCIENCE2002} Z. Tang, N.~A. Kotov, and M. Giersig, Science \textbf{297}, 237 (2002).
\bibitem{katzPRL2002} D.\ Katz, T.\ Wizansky, O.\ Millo, E.\ Rothenberg, T.\ Mokari, and U.\ Banin, Phys.\ Rev.\ Lett.\ {\bf 89}, 086801 (2002); {\em ibid.} 199901 (2002)
\bibitem{johnsonNATMAT2002} J.\ C.\ Johnson, H.\ J.\ Choi, K.\ P.\ Knutsen, R.\ D.\ Schaller, P.\ D.\ Yang, R.\ J.\ Saykally, Nat. Mater.\ {\bf 1}, 106 (2002).
\bibitem{bjorkNANOLETT2002} M.\ T.\ Bj\"{o}rk, B.\ J.\ Ohlsson, T.\ Sass, A.\ I.\ Persson, C.\ Thelander, M.\ H.\ Magnusson, K.\ Deppert, L.\ R.\ Wallenberg, and L.\ Samuelson, Nano Lett.\ {\bf 2}, 87 (2002). 
\bibitem{bjorkAPL2002} M.\ T.\ Bj\"{o}rk, B.\ J.\ Ohlsson, C.\ Thelander, A.\ I.\ Persson, K.\ Deppert, L.\ R.\ Wallenberg, and L.\ Samuelson, Appl.\ Phys.\ Lett.\ {\bf 81}, 4458 (2002).
\bibitem{gudiksenNATURE2002} M.\ S.\ Gudiksen, L.\ J.\ Lauhon, J.-F.\ Wang, D.\ C.\ Smith, and C.\ M.\ Lieber, Nature {\bf 415}, 617 (2002).
\bibitem{krishnamachariAPL2004} U.\ Krishnamachari, M.\ Borgstr\"{o}m, B.\ J.\ Ohlsson, N.\ Panev, L.\ Samuelson, W.\ Seifert, M.\ W.\ Larsson and L.\ R.\ Wallenberg, Appl.\ Phys.\ Lett.\ {\bf 85}, 2077 (2004).
\bibitem{samuelsonPHYSICAE2004} L.\ Samuelson, M.\ T.\ Bj\"{o}rk, K.\ Deppert, M.\ Larsson, B.\ J.\ Ohlsson, N.\ Panev, A.\ I.\ Persson, N.\ Sk\"{o}ld, C.\ Thelander and L.\ R.\ Wallenberg, Physica E {\bf 21}, 560 (2004)
\bibitem{huangSMALL2005} Y.\ Huang, X.\ Duan and C.\ M.\ Lieber, Small {\bf 1}, 142 (2005).
\bibitem{thelanderNANOLETT2005} C.\ Thelander, H.\ A.\ Nilsson, L.\ E.\ Jensen and L.\ Samuelson, Nano Lett.\ {\bf 5}, 635 (2005).
\bibitem{yangSCIENCE2005} C.\ Yang, Z.\ Zhong, C.\ M.\ Lieber, Science {\bf 310}, 1304 (2005).
\bibitem{luJPD2006} W.\ Lu, C.\ M.\ Lieber, J.\ Phys.\ D {\bf 39},  R387 (2006).
\bibitem{pauzauskieMATERIALSTODAY2006} P.\ Pauzauskie, P.\ Yang, Materials Today {\bf 9}, 36 (2006).
\bibitem{perssonNANOLETT2006} A.\ I.\ Persson, M.\ T.\ Bj\"{o}rk, S.\ Jeppesen, J.\ B.\ Wagner, L.\ R.\ Wallenberg and L.\ Samuelson, Nano Lett.\ {\bf 6}, 403 (2006).
\bibitem{samuelsonNANOTECH2006} L.\ Samuelson, Nanotechnology {\bf 17}, (2006).
\bibitem{johanssonNATMAT2006} J.\ Johansson, L.\ S.\ Karlsson, C.\ P.\ T.\ Svensson, T.\ M\aa rtensson, B.\ A.\ Wacaser, K.\ Deppert, L.\ Samuelson and W.\ Seifert, Nat. Mater.\ {\bf 5}, 574 (2006). 
\bibitem{janikAPL2006} E.\ Janik, J.\ Sadowski, P.\ D\l u\`{z}ewski, S.\ Kret, L.\ T.\ Baczewski, A.\ Petroutchik, E.\ \L usakowska, J.\ Wrobel, W.\ Zaleszczyk, G.\ Karczweski, and T.\ Wojtowicz, Appl.\ Phys.\ Lett.\ {\bf 89}, 133114 (2006).
\bibitem{thelanderMATERIALSTODAY2006} C.\ Thelander, P.\ Agarwal, S.\ Brongersma, J.\ Eymery, L.\ F.\ Feiner, A.\ Forchel, M.\ Scheffler, W.\ Riess, B.\ J.\ Ohlsson, U.\ G\"{o}sele and L.\ Samuelson, Materials Today {\bf 9}, 28 (2006).
\bibitem{martelliNANOLETT2006} F.\ Martelli, S.\ Rubini, M.\ Piccin, G.\ Bais, F.\ Jabeen, S.\ De Franceschi, V.\ Grillo, E.\ Carlino, F.\ D'Acapito, F.\ Boscherini, S.\ Cabrini, M.\ Lazzarino, L.\ Businaro, F.\ Romanato, and A.\ Franciosi, Nano Lett.\ {\bf 6}, 2130 (2006).
\bibitem{lieberNATURE2006} J. Xiang, W. Lu, Y. Hu, Y. Wu, H. Yan, and C. M. Lieber, 
Nature {\bf 441}, 489 (2006).
\bibitem{fengNANOLETT2007} X.\ Feng, R.\ He, P.\ Yang, M.\ Roukes, Nano.\ Lett.\ {\bf 7}, 1953 (2007).
\bibitem{NeretinaNANOTECH2007} S. Neretina, R.~A. Hughes, J.~F. Britten, N.~V. Sochinskii, J.~S. Preston, and P. Mascher, Nanotechnology \textbf{18}, 275301
(2007).
\bibitem{liNANOLETT2007} H.-Y.\ Li, O.\ Wunnicke, M.\ T.\ Borgstr\"{o}m, W.\ G.\ G.\ Immink, M.\ H.\ M.\ van Weert, M.\ A.\ Verheijen, and E.\ P.\ A.\ M.\ Bakkers, Nano Lett.\ {\bf 7}, 1144 (2007).
\bibitem{zhuNANOLETT2007} Q.\ Zhu, K.\ F.\ Karlsson, E.\ Pelucchi, and E.\ Kapon, Nano Lett.\ {\bf 7}, 2227 (2007).
\bibitem{hochbaumNATURE2008} A.\ I.\ Hochbaum, R.\ Chen, R.\ D.\ Delgado, W.\ Liang, E.\ C.\ Garnett, M.\ Najarian, A.\ Majumdar, P.\ Yang, Nature {\bf 451}, 163 (2008).
\bibitem{sorensenAPL2008} B.\ S.\ S\o{}rensen, M.\ Aagesen, C.\ B.\ S\o{}rensen, P.\ E.\ Lindelof, K.\ L.\ Martinez, and J.\ Nyg\aa{}rd, Appl.\ Phys.\ Lett.\ {\bf 92}, 012119 (2008).
\bibitem{jagadishNANOTECH2008} M.~S. Song, J.~H. Jung, Y. Kim, Y. Wang, J. Zou, 
H.~J. Joyee, Q. Gao, H.~H. Tan, and C. Jagadish, Nanotechnology {\bf 19}, 125602
(2008).
\bibitem{roddaroPRL2008} S. Roddaro, A. Fuhrer, P. Brusheim, C. Fasth, H.~Q. Xu,
L. Samuelson, J. Xiang, and C. M. Lieber, Phys. Rev. Lett. {\bf 101}, 186802 (2008).
\bibitem{wojtowiczNANOLETT2008}  W.\ Zaleszczyk,  E.\ Janik,  A.\ Presz,  P.\ 
D{\l}u{\.z}ewski,  S.\ Kret,  W.\ Szuszkiewicz,  J.-F.\ Morhange,  E.\ Dynowska,  H.\ 
Kirmse,  W.\ Neumann,  A.\ Petroutchik,  L.\ T.\ Baczewski,  G.\ Karczewski,  and
T.\ Wojtowicz, Nano Lett. \textbf{8}, 4061(2008).

\bibitem{wolfSCIENCE2001} S.\ A.\ Wolf, D.\ D.\ Awschalom, R.\ A.\ Buhrmann, J.\ M.\ Daughton, S.\ von Moln\'{a}r, M.\ L.\ Roukes, A.\ Y.\ Chtchelkanova, and D.\ M.\ Treger, Science {\bf 294}, 1488 (2001).
\bibitem{zuticRMP2004} I.\ Zuti\'{c}, J.\ Fabian, and S.\ Das Sarma, Rev.\ Mod.\ Phys.\ {\bf 76}, 323 (2004).
\bibitem{molenkampNATURE1999} R.\ Fiederling, M.\ Keim, G.\ Reuscher, W.\ 
Ossau, G.\ Schmidt, A.\ Waag, and L.\ W.\ Molenkamp, Nature \textbf{402}, 787 
(1999).
\bibitem{ohnoNATURE1999} Y.\ Ohno, D.\ K.\ Young, B.\ Beschoten, F.\ Matsukura,
H.\ Ohno, and D.\ D.\ Awschalom, Nature \textbf{402}, 790 (1999).
\bibitem{bychkovJPC1984} Y.\ A.\ Bychkov and E.\ I.\ Rashba, J.\ Phys.\ C {\bf 17}, 6039 (1984).
\bibitem{dattaAPL1990} S.\ Datta and B.\ Das, Appl.\ Phys.\ Lett.\ {\bf 56}, 665 (1990).

\bibitem{winklerbook} R.\ Winkler, {\it Spin-Orbit Coupling Effects in Two-Dimensional Electron and Hole Systems} (Springer, Berlin, 2003).
\bibitem{winklerPRL2000} R.\ Winkler, S.\ J.\ Papadakis, E.\ P.\ De Poortere, and M.\ Shayegan, Phys.\ Rev.\ Lett.\ {\bf 85}, 4574 (2000).
\bibitem{danneauPRL2006} R.\ Danneau, O.\ Klochlan, W.\ R.\ Clarke, L.\ H.\ Ho, A.\ P.\ Micolich, A.\ R.\ Hamilton, M.\ Y.\ Simmons, M.\ Pepper, D.\ Ritchie, and U.\ Z\"{u}licke, Phys.\ Rev.\ Lett.\ {\bf 97}, 026403 (2006).
\bibitem{koduvayurPRL2008} S.\ P.\ Koduyavur, L.\ P.\ Rokhinson, D.\ C.\ Tsui, L.\ N.\ Pfeiffer, and K.\ W.\ West, Phys.\ Rev.\ Lett.\ {\bf 100}, 126401 (2008).
\bibitem{pradoPRB2004} S.\ J.\ Prado, C.\ Trallero-Giner, A.\ M.\ Alcalde, V.\ L\'{o}pez-Richard, and G.\ E.\ Marques, Phys.\ Rev.\ B {\bf 69}, 201310 (2004).
\bibitem{pryorPRL2006} C.\ E.\ Pryor, and M.\ E.\ Flatt\'e, Phys.\ Rev.\ Lett.\ {\bf 96}, 026804 (2006). 
\bibitem{zhangAPL2007} X.-W.\ Zhang, W.-J.\ Fan, K.\ Chang, S.-S.\ Liu, and J.-B.\ Xia, Appl.\ Phys.\ Lett.\ {\bf 91}, 113108 (2007).
\bibitem{haendelPRL2006} K.-M.\ Haendel, R.\ Winkler, U.\ Denker, O.\ G.\ Schmidth, and R.\ J.\ Haug, Phys.\ Rev.\ Lett.\ {\bf 96}, 086403 (2006).
\bibitem{luttingerPR1956} J.\ M.\ Luttinger, Phys.\ Rev.\ {\bf 102}, 1030 (1956).
\bibitem{suzukiPRB1974} K.\ Suzuki and J.\ C.\ Hensel, Phys. Rev. B \textbf{9},
4184 (1974).
\bibitem{shermanPLA88} E.\ I.\ Rashba and E.\ Yu.\ Sherman, Phys. Lett. A
\textbf{129}, 175 (1988).

\bibitem{luPRB2006} C.\ L\"{u}, J.\ L.\ Cheng, and M.\ W.\ Wu, Phys.\ Rev.\ B {\bf 73}, 125314 (2006).
\bibitem{bastardrev} G.\ Bastard, J.\ A.\ Brum, and R.\ Ferreira, {\it Solid State Physics} (Academic Press, San Diego, 1991), vol.\ 44, pp.\ 229-415.
\bibitem{perssonNANOLETT2004} M.\ P.\ Persson and H.\ Q.\ Xu, Nano Lett.
\textbf{4}, 2409 (2004); Phys. Rev. B \textbf{73}, 125346 (2006).
\bibitem{zhangEPJB2006} X.~W. Zhang, Y.~H. Zhu, and J.~B. Xia. Eur. Phys. J. B
{\bf 52}, 133 (2006).
\bibitem{haradaPRB2006} Y.\ Harada, T.\ Kita and O.\ Wada, Phys.\ Rev.\ B {\bf 74}, 245323 (2006).
\bibitem{csontosPRB2007} D.\ Csontos and U.\ Z\"{u}licke, Phys.\ Rev.\ B. \textbf{76}, 073313 (2007).
\bibitem{kyrchenkoPRB2004} F.\ V.\ Kyrchenko, and J.\ Kossut, Phys.\ Rev.\ B {\bf 70}, 205317 (2004).
\bibitem{lewyanvoonNANOLETT2004} L.\ C.\ Lew Yan Voon, R.\ Melnik, B.\ Lassen, M.\ Willatzen, Nano Lett.\ {\bf 4}, 289 (2004).
\bibitem{shengPRB2007} W.\ Sheng, and A.\ Babinski, Phys.\ Rev.\ B {\bf 75}, 033316 (2007).
\bibitem{troncaleAPL2007} V.\ Troncale, K.\ F.\ Karlsson, E.\ Pelucchi, A. Rudra, and E.\ Kapon, Appl.\ Phys.\ Lett.\ {\bf 91}, 241909 (2007).
\bibitem{csontosAPL2008} D.\ Csontos and U.\ Z\"{u}licke, Appl.\ Phys.\ Lett.\  {\bf 92}, 023108 (2008).
\bibitem{csontosPRB2008} D. Csontos, U. Z\"ulicke, P. Brusheim, and H.\ Q.\ Xu,
Phys. Rev. B \textbf{78}, 03307 (2008).
\bibitem{chenPRB2005} P.\ Chen, Phys.\ Rev.\ B {\bf 72}, 045335 (2005).
\bibitem{winklerPRB2004} R.\ Winkler, Phys.\ Rev.\ B {\bf 70}, 125301 (2004).
\bibitem{ngNANOLETT2004} H.\ T.\ Ng, J.\ Han, T.\ Yamada, P.\ Nguyen, Y.\ P.\ Chen, M.\ Meyyappan, Nano Lett.\ {\bf 4}, 1247 (2004).
\bibitem{bryllertEDL2006} T.\ Bryllert, L.\ E.\ Wernersson, L.\ E.\ Froberg, L.\ Samuelson, IEEE Electron Dev.\ Lett.\ {\bf 27}, 323 (2006).
\bibitem{wangAPL2000} Q.\ Wang, N.\ Carlsson, I.\ Maximov, P.\ Omling, L.\ Samuelson, W.\ Seifert, W.\ Sheng, I.\ Shorubalko, and H.\ Q.\ Xu, Appl.\ Phys.\ Lett.\ {\bf 76}, 2274 (2000).
\bibitem{danneauAPL2006} R.\ Danneau, W.\ R.\ Clarke, O.\ Klochan, A.\ P.\ Micolich, 
A.\ R.\ Hamilton, M.\ Y.\ Simmons, M.\ Pepper, and D.\ A.\ Ritchie, Appl. Phys. Lett.
\textbf{88}, 012107 (2006).
\bibitem{klochanAPL2006} O.\ Klochan, W.\ R.\ Clarke, R.\ Danneau, A.\ P.\
Micolich, L.\ H.\ Ho, A.\ R.\ Hamilton, K.\ Muraki, and Y.\ Hirayama, Appl. Phys. Lett.
\textbf{89}, 092105 (2006).
\bibitem{ensslinAIPPROC2007} B.~Grbi\'c, R.\ Leturcq, T.\ Ihn, K.\ Ensslin,
D.\ Reuter, and A.~D.\ Wieck, AIP Conf. Proc. {\bf 893}, 777 (2007).

\bibitem{baldereschiPRB1973} A.\ Baldereschi and N.\ O.\ Lipari, Phys.\ Rev.\ B {\bf 8}, 2697 (1973).
\bibitem{baldereschiPRB1974} A.\ Baldereschi and N.\ O.\ Lipari, Phys.\ Rev.\ B {\bf 9}, 1525 (1974)
\bibitem{sercelAPL1990} P.~C.\ Sercel and K.~J.\ Vahala, Appl.\ Phys.\ Lett.\ {\bf 57}, 545 (1990).
\bibitem{sercelPRB1990} P.~C.\ Sercel and K.~J.\ Vahala, Phys.\ Rev.\ B {\bf 42}, 3690 (1990).
\bibitem{vurgaftmanJAP2001} I.\ Vurgaftman, J.~R.\ Meyer, and L.~R.\ Ram-Mohan, J.\ Appl.\ Phys.\ {\bf 89}, 5815 (2001).
\bibitem{perssonAPL2002} M.\ P.\ Persson and H.\ Q.\ Xu, Appl.\ Phys.\ Lett.\ {\bf 81}, 1309 (2002).
\bibitem{dietlPRB2001} T.\ Dietl, H.\ Ohno, and F.\ Matsukura, Phys.\ Rev.\ B {\bf 63}, 195205 (2001).

\bibitem{vweesPRL1988} B.\ J.\ van Wees, H.\ van Houten, C.\ W.\ J.\ Beenaker, J.\ G.\ Williamson, L.\ P.\ Kouwenhoven, D.\ van der Marel, and C.\ T.\ Foxon, Phys.\ Rev.\ Lett.\ {\bf 60}, 848 (1988).
\bibitem{wharamJPC1988} D.\ A.\ Wharam, T.\ J.\ Thornton, R.\ Newbury, M.\ Pepper, H.\ Ahmed, J.\ E.\ F.\ Frost, D.\ G.\ Hasko, D.\ C.\ Peacock, D.\ A.\ Ritchie, and G.\ A.\ C.\ Jones, J.\ Phys.\ C {\bf 21}, L209 (1988).

\bibitem{zhangJPD2007} X.~W. Zhang and J.~B. Xia, J. Phys. D {\bf 40}, 541 (2007).

\end{thebibliography}
\end{document}